\documentclass[a4paper,12pt]{article}
\usepackage[utf8x]{inputenc}
\usepackage{amssymb}
\usepackage{lscape,amsmath,amsxtra}
\usepackage{latexsym}
\usepackage{bm}
\usepackage{amsfonts}
\usepackage{natbib}
\usepackage[pdftex]{graphicx}

\def\E{\mathbb{E}}
\def\Var{\mathbb{V}ar}
\def\E{\mathbb{E}}
\def\Cov{\mathbb{C}ov}

\title{Estimation of spatial  max-stable models using   threshold exceedances}
\author{Jean-Noel Bacro \\
I3M, Universit\'e Montpellier II\\
and\\
Carlo Gaetan\\
DAIS,
Universit\`a Ca' Foscari - Venezia }

\date{}
\begin{document}

\maketitle
\begin{abstract}
 Parametric inference for spatial max-stable 
processes  is difficult  
since the related likelihoods are unavailable.  A composite likelihood approach based on the bivariate distribution of  block maxima  has 
been recently proposed in the literature. However modeling  block maxima is a wasteful approach provided that other information is available. 
Moreover an approach based on block, typically annual, maxima is unable to take into account the fact that maxima occur or not 
simultaneously.  If  time series of, say, daily data are available, then estimation procedures  based on  exceedances of a  high  threshold could mitigate such problems. 
In this paper we focus on two approaches for composing likelihoods based on 
pairs of exceedances. The first one comes from the tail approximation for 
bivariate distribution proposed by \citet{Ledford:Tawn:1996} when both pairs of observations  exceed the fixed threshold. The second one 
uses the bivariate extension \citep{Rootzen:Tajvidi:2006} of the generalized Pareto distribution  which allows to model exceedances when at least one of the 
components is over the threshold. The two approaches are compared through a simulation study according to different degrees of spatial dependency. Results show  that  both the  strength of the spatial dependencies  and the threshold choice play a 
fundamental role in determining which is the best estimating  procedure.
\end{abstract}
\noindent \textit{AMS 2000 subject classification}: primary 62G32, secondary  62M30.

\noindent Keywords: Composite likelihood, Extremal dependence, Generalized Pareto distribution, Spatial statistics.

\section{Introduction}   
Extreme value theory for multivariate data is well established 
and there are a lot
of results which can be used to characterize extreme values
distributions
\citep[see, for instance,][]{Resnick:1987,Beirlant:Goegebeur:Segers:Teugels:2004}. 

From a
statistical point of view \citep{Coles:2001} there are mainly two approaches to fitting
multivariate models: block maxima and threshold exceedances. Block
maxima is the most used approach: assuming a 
parametric family of distributions, the model parameters are estimated on a block
maxima sample. In practice,   bivariate distributions of maxima have been frequently considered
and there are few papers that deal  with higher
dimensions \citep{Tawn:1988,Tawn:1990, Coles:Tawn:1991,
  Coles:Tawn:1994}.
  
  Recently, spatial extreme problems have received
an increasing attention in the literature (see, among others, \citet{Coles:1993, Coles:Tawn:1996,
   Casson:Coles:1999, Buishand:de_Haan:Zhou:2008, Bel:Bacro:Lantuejoul:2008,
  Padoan:Ribatet:Sisson:2010} and for  recent reviews  \citet{Bacro:Gaetan:2012,
  Davison:Padoan:Ribatet:2012}). 
In particular max-stable processes
\citep{De_Haan:1984} take a prominent role   in modeling  the spatial dependence   because they are a natural extension of the
multivariate extreme value distributions for dealing with  spatial data.

However, in spite of the relative large number of instances \citep{Smith:1990, Schlather:2002,
  Kabluchko:Schlather:de_Haan:2009},   parametric likelihood inference for such models is
not easy because the related likelihood is unavailable or too
cumbersome to compute.
For this reason
 a methodology based on partial specification of
the full likelihood, the composite likelihood
\citep{Lindsay:1988, Varin:Reid:Firth:2011}, has  gained popularity.
For spatial extreme data, 
the composite likelihood \citep{Smith:Stephenson:2009,Padoan:Ribatet:Sisson:2010} is  built by taking product of
pairwise distributions. Then parameter estimates are  obtained by  fitting the
model to block bivariate maxima vectors. 

In this paper we are concerned
with estimation procedures based on exceedances of a large threshold. 
In doing this we expect that the efficiency of the estimates
can be    improved in all situations where 
the paucity of data prevents the traditional approach based on the block maxima.
For instance,  rainfall data are frequently collected  every day
but for few  years.  If annual maxima are considered, 
we have got  estimates  by using few  repetitions.
A  larger data set can be  obtained by 
dealing with daily exceedances of a large threshold,
even if 
constraints of stationarity can reduce the temporal window  to some months per years.

Our aim is not suggesting new models for explaining the spatial behaviour of exceedances
\citep[see, for instance,][]{Buishand:de_Haan:Zhou:2008,Turkman:Turkman:Pereira:2010} but proposing and comparing 
two different approaches for composing  likelihoods based on exceedances.
The approaches rely on  different specifications of the bivariate distribution
of the  exceedances.

The first one comes from
\citet{Ledford:Tawn:1996}. These authors have proposed a tail approximation
for the bivariate distribution valid when both components exceed a large
threshold.  
Since the  approximation is  valid for
values simultaneously larger than the threshold, a censored
version of this approximation is proposed for pairs of data having one or two
components below the threshold. 

The second approach we have considered is based on a
result for multivariate exceedances due to 
\citet{Rootzen:Tajvidi:2006}. This result allows us
to model exceedances over a threshold  using a multivariate  extension of the 
generalized Pareto distribution when at least one of the components is over
the threshold. The result is asymptotic and leads to 
approximating  the conditional distribution
given that  at least one component is over the threshold. 

The paper is organized as follows. Spatial max-stable processes are
presented in Section \ref{sec:max-stable}. The two bivariate  models for the exceedances
 are described in Section  \ref{sec:exceedances} and  
the estimation  procedure
we propose  is
detailed in section \ref{sec:inference}.
Section \ref{sec:simulation} presents a  simulation study where the two threshold exceedances models are compared, according different degrees of spatial extreme dependencies. 
Finally, some
perspectives are discussed in Section \ref{sec:discussion}.

\section{Spatial max-stable processes}\label{sec:max-stable}
When we deal with extreme values, the traditional approach considers
the maxima over a number of replications using the results
from the extreme value theory.
This theory embraces the univariate and the multivariate case \citep[see, for instance, ][for a recent account]
{Beirlant:Goegebeur:Segers:Teugels:2004}.
In the case of spatial data  maxima of observations over a fixed  period
are frequently collected on a finite subset  of sites 
and the aim  is to characterize the stochastic  behavior 
on an uncountable  set of unsampled sites.
For  this,  multivariate models are unsuitable and their natural extensions 
 are the  max-stable processes \citep{De_Haan:1984}.

Let $\{Z(s), s\in {\cal S}\}$ be a stochastic process on an index set
$ {\cal S}\subset \mathbb{R}^d$. 
We assume that $Z_k(\cdot), 1\leq k\leq n$, are $n$ independent copies of  $Z(\cdot)$.
The process $Z(\cdot)$ is a (spatial) max-stable process if 
the multivariate distribution of $p\geq 1$ sites $Z(s_1),\ldots,Z(s_p)$, with  $s_i\in {\cal S}$ and $i=1,\ldots,p$,
satisfies the so-called max-stability property, i.e. 
there exist sequences $a_{n}(s)>0$, $b_{n}(s)$, 
   $n\geq 1$, such that
\begin{equation}
\label{eq:maxstable1}
\Pr\left(\max_{1\leq k \leq n} \frac{ Z_k(s_j)- b_n(s_j)}{a_n(s_j)}\le z_j;\,{1\leq j\leq p}\right) =
\Pr\left(Z(s_j) \le z_j;\,{1\leq j\leq p}\right)
\end{equation}

From the definition,  any finite distribution of a max-stable process is an  extreme value  distribution. In the sequel we will denote
the $p$ dimensional finite  distribution of a spatial max-stable process
as $G_{s_1,\ldots,s_p}(z_1,\ldots,z_p)$.

If  we  suppose that we have  $n$ independent copies of a process $Y(\cdot)$ such that
\begin{equation}
\label{eq:maxstable2}
\lim_{n\rightarrow\infty}\Pr\left(\max_{1\leq k \leq n} \frac{ Y_k(s_j)- \beta_n(s_j)}{\alpha_n(s_j)}\le z_j;\,{1\leq j\leq p}\right) =
G_{s_1,\ldots,s_p}(z_1,\ldots,z_p)
\end{equation}
for suitable sequences $\alpha_n(s) > 0$, $\beta_n(s)$, we say that the distribution
of $Y(s_1),\ldots,Y(s_p)$, i.e.  $F_{s_1,\ldots,s_p}$, is in the attraction domain of the   extreme value distribution $G_{s_1,\ldots, s_p}$. This fact will be denoted by $F_{s_1,\ldots,s_p}\in \mathcal{D}(G_{s_1,\ldots,s_p})$.
 
There are several  representations for max-stable processes, see for instance
\citet{Smith:1990,Schlather:2002,Kabluchko:Schlather:de_Haan:2009,Lantuejoul:Bacro:Bel:2010}. 
Just for illustration,   in the sequel we will concentrate on the 
 Gaussian
extreme value process \citep{Smith:1990}.

Let $\Pi$ be a Poisson process of intensity
$x^{-2}dxds$ on $(0,\infty)\times \mathcal{S}$.  Then,
\begin{equation}
\label{eqSmith}
Z(s)={ \max_{(x,u)\in \Pi}}\,x f(s-u),
\end{equation}
defines a stationary max-stable process, with unit Fr\'echet marginal
distribution, i.e.  $\Pr(Z(s)\le z)=\exp(-1/z)$, $z>0$, provided that
 $f(\cdot)$ is a non negative function such that $\int_\mathcal{S} f(s)ds=1$.
Choosing $f$ as
the bivariate Gaussian density 
$$f({ s})=(2\pi)^{-1}\mid\Sigma\mid^{-1/2}\exp\left(
  -\frac{1}{2}{ s}'\Sigma^{-1} { s}\right), 
$$
where  $\Sigma$ is a covariance matrix,
 leads to the  Gaussian
extreme value process.
\citet{Smith:1990} gives an intuitive physical interpretation of the process
 (\ref{eqSmith})
 in terms of rainfall-storms
i.e. $x$ and $u$ are, respectively,  the size and the center of a  'storm' that
spreads out according the density $f$.

For max-stable processes analytical expressions for multivariate distributions
are quite cumbersome 
and  explicit formulas are known mainly for the bivariate  case.
 \cite{Smith:1990}  derived the bivariate distribution of (\ref{eqSmith})   for two different sites $s_i$ and $s_j$, namely
\begin{equation}\label{eq:biv_dist_max-stable}
G_{s_i,s_j}(z_i,z_j)=\exp\left\{ -\frac{1}{z_i}\Phi\left(\frac{a_{ij}}{2}+\frac{1}{a_{ij}}\log
    \frac{z_j}{z_i}\right) - \frac{1}{z_j}\Phi\left(\frac{a_{ij}}{2}+\frac{1}{a_{ij}}\log
    \frac{z_i}{z_j}\right)\right\}
    \end{equation}
where $\Phi(\cdot)$ denotes the standard normal cumulative distribution
function and $a_{ij}=\sqrt{(s_i-s_j)'\Sigma^{-1}(s_i-s_j)}$.
In the case of a marginal Gumbel distribution, i.e. $\Pr(Z(s) \le z)=\exp\{-\exp (-z)\}$, the bivariate distribution is
given by
$$  G_{s_i,s_j}(z_i,z_j) =  \exp\left\{
-e^{-z_i} \Phi\left( \frac{a_{ij}}{2} +
    \frac{z_i-z_j}{a_{ij}}\right) 
- e^{-z_j} \Phi\left(
    \frac{a_{ij}}{2} + \frac{
      z_j-z_i}{a_{ij}}\right)
\right\}
$$

More recently \citet{Genton:Ma:Sang:2011} have derived multivariate expression for the
Gaussian extreme value process
but in practice only the three-variate case is simple  to deal with.

This fact has restricted the application of the parametric inference based on the full likelihood and motivated approaches based on the composite likelihood
\citep{Lindsay:1988}.
Roughly speaking (see section  \ref{sec:inference} for more details) the composite likelihood
is an estimating function obtained combining  marginal or conditional densities. 

However, these new advances still require a lot of temporal information, 
because we need to calculate the maxima over several period
for providing enough sample data.
This weakness can be overcome if   we use all observations that exceed some fixed (high) threshold.
For this reason in the next section we will illustrate  two different ways for modeling
bivariate threshold exceedances and we will exploit these representations for
building suitable composite likelihoods.

 \section{Models for bivariate threshold exceedances}\label{sec:exceedances}

 Univariate extreme values results \citep{Pickands:1975,Davison:Smith:1990} justify to model the conditional distribution of the excesses $Y(s)-u$ through 
the Generalized Pareto (GP) distribution
\begin{displaymath}
\Pr\left(\frac{Y(s)-u}{\sigma_u}\le z | Y(s) > u\right)\approx 1-\left(1+\xi {z}\right)^{-1/\xi},\qquad z \ge 0,
\end{displaymath} 
provided that
 $F_s(\cdot)\in \mathcal{D}(G)$.
Here  $\xi$ is a real shape parameter  and $\sigma_u$ is a positive scale parameter, which depends on the threshold $u$.

 The GP distribution provides an
approximation  of the tail of the  distribution function in the sense  that
for a large enough value of $u$, we have
\begin{equation}
\label{eq1Coles}
\Pr\left(\frac{Y(s)-u}{\sigma_u}\le z  \right) \approx 1-\zeta\left\{1+\xi{ z}\right\}^{-1/\xi},
\qquad z \ge 0,
\end{equation}
where $\zeta=1-F_s(u)$.

Now we consider the bivariate  random variable $Y(s_i),Y(s_j)$  with the same marginal distributions and
$F_{s_i,s_j}\in\mathcal{D}(G_{s_i,s_j})$.
According (\ref{eq1Coles}), the integral transformation
 \begin{equation}\label{eq:it}
  \tilde{Y}_k=-\left( \log \left\{1-\zeta\left[
       1+\xi\frac{{Y}(s_k)-u}{\sigma_u}\right]^{-1/\xi}\right\}\right)^{-1}, \, k=i,j
\end{equation}
has unit Fr\'echet distribution.

The bivariate  distribution $F_{s_i,s_j}(y_i,y_j)$ can
 be approximated  on  $(u,\infty)^2$ \citep{Ledford:Tawn:1996}  
 as 
 \begin{equation}\label{eq:lt}
 F_{s_i,s_j}(y_i,y_j)\approx G_{s_i,s_j}(\tilde{y}_i,\tilde{y}_j).
 \end{equation}

Formula (\ref{eq:lt})  gives us an approximation of the bivariate distribution in the tail when
both components are greater than threshold $u$.

Another approach  \citep{, Beirlant:Goegebeur:Segers:Teugels:2004,Rootzen:Tajvidi:2006}
considers a bivariate distribution of large values, 
given that at least one of the components is large.
This distribution is   a bivariate
analogue for the GP distribution.

Provided that 
$F_{s_i,s_j}
(\cdot,\cdot)\in {\cal D}(G_{s_i,s_j}(\cdot,\cdot))$
and   $0<G_{s_i,s_j}({0},0)<1$,
\citet{Rootzen:Tajvidi:2006}   show that 
for large $u$
$$\Pr\left(
  \frac{Y(s_i)-u}{\sigma_u } \leq z_i, \frac{Y(s_j)-u}{\sigma_u } \leq z_j
  \mid
  Y(s_i) > u  \mbox{ or } Y(s_j) > u\right)
\approx
H_{s_i,s_j}({z_i,z_j}),
$$
where 
\begin{equation}\label{eq:rt}
 H(x_i,x_j)=\frac{1}{-\log(G_{s_i,s_j}({0},0))}
 \log \frac{G_{s_i,s_j}(x_i,x_j)}{G_{s_i,s_j}(\min(x_i,0),\min(x_j,0))}.
\end{equation}

 Actually the Rootzen-Tajvidi's result applies to a dimension greater than two, and in the sequel we refer to the distribution (\ref{eq:rt}) as the bivariate GP distribution.

\section{Inference procedures for spatial high threshold excesses}
\label{sec:inference}
As mentioned in section \ref{sec:max-stable}, likelihood inference for spatial
max-stable processes is difficult because we don't known an analytical
expression for the multivariate distribution of the majority of the
spatial max-stable processes.

The composite likelihood (CL)  \citep{Lindsay:1988} is an inference function built by multiplying 
the likelihood of marginal or
conditional events.
More precisely, let $\{f(\cdot,\theta ), \theta\in {\Theta}\}$ be a parametric family of joint densities for the  observations  $y_1,\ldots,y_n\in D\subseteq \mathbb{R}^n$
and
consider a set of events $\{A_i:A_i\subseteq \mathfrak{F}, i\in I \subseteq \mathbb{N}\}$,  where $ \mathfrak{F}$ is a $\sigma$-algebra on $D$.
 The  logarithm of the CL is defined as \citep{Lindsay:1988}:
$$
cl(\theta)=\sum_ {i\in I} w_i\log f((y_1,\ldots,y_n)'\in
A_i,\theta ),
$$
where $w_i$ are non negative weights to be fixed.

For max-stable
processes, \citet{Padoan:Ribatet:Sisson:2010} have proposed a CL
approach based on the pairwise distributions for estimating the parameters of the distribution of  random vectors of  block maxima. Their
estimation  procedure has been implemented in the R package
  \texttt{SpatialExtremes} 
\citep{SpatialExtremes} 
available on the CRAN repositories.
Here we borrow the same idea but we use exceedances instead of block 
maxima.

We assume that data $\{y_{t1},\ldots,y_{tp}\}$, $t=1,\ldots,T$, are $T$ independent realizations
  of the vector $\{Y(s_1),\ldots,Y(s_p)\}$, where  $s_1,\ldots,s_p$ are
   $p$ sites belonging to $\mathcal{S}$.

Let $ f_{s_i,s_j} (\cdot,\cdot;\theta)$ be the bivariate density  associated to the distribution
(\ref{eq:lt}) or (\ref{eq:rt}).
The CL estimate $\hat{\theta}_T$ is obtained maximizing 
 \begin{equation}
 \label{CLvrais}
 cl_T(\theta)=\sum_{t=1}^T\sum_{i=1}^p \sum_{j >  i}^p  w_{ij} \log
 f_{s_i,s_j}(y_{ti},y_{tj};\theta)=\sum_{t=1}^TU_t(\theta)
 \end{equation}
where  $w_{ij}\ge 0$ designate user-specified weights \citep{Padoan:Ribatet:Sisson:2010}
 and $U_t(\theta)=\sum_{i=1}^p \sum_{j >  i}^p  w_{ij} \log
 f_{s_i,s_j}(Y_{ti},Y_{tj};\theta)$.

Note that there is a fundamental difference between the  estimating functions,
when we choose  (\ref{eq:lt}) or (\ref{eq:rt}). 
In using (\ref{eq:lt}) we  exploit a tail approximation, moreover  the approximation  is sound on the region $(u,\infty)^2$ and does not apply to  pairs of observations outside that region. For pairs such that at least one component does not exceed $u$, we use  the censored approach  as proposed in \citet{Ledford:Tawn:1996}.
The likelihood 
contribution of each pair is 
$$f_{s_i,s_j}({y}_{ti},{y}_{tj};\theta)=\left\{
\begin{array}{ll}
\nabla_{ab} G_{s_i,s_j}(\tilde{y}_{ti},\tilde{y}_{tj};\theta) & \mbox{\rm
  if } y_{ti} \geq u, y_{tj}\geq u\\[0.2cm]
\nabla_{a}G_{s_i,s_j}(\tilde{y}_{ti},\tilde{u};\theta)
& \mbox{\rm
  if }  y_{ti} \geq u, y_{tj} < u \\[0.2cm]
  \nabla_{b}G_{s_i,s_j}(\tilde{u},\tilde{y}_{tj};\theta)
& \mbox{\rm
  if }  y_{ti}  < u, y_{tj} \ge u \\[0.2cm]
G_{s_i,s_j}(\tilde{u},\tilde{u};\theta) & \mbox{\rm
  if }  y_{ti}  < u, y_{tj} < u\\[0.2cm]
\end{array}
\right.
$$
where $\theta=(\sigma_u,\xi,\beta)$ and $\beta$ is the (possibly) vector of spatial dependence parameters.
Instead of maximizing $cl(\theta)$ with respect to  all  parameters,
we can estimate $\theta$  in two steps.
First of all we estimate the marginal
parameters $\xi$ et $\sigma_u$, then their estimates, 
$\hat{\xi}$ and $\hat{\sigma_u}$, are plugged in 
(\ref{eq:it}) for obtaining an estimate for  $\beta$ maximizing
$cl(\hat{\xi},\hat{\sigma_u},\beta)$. 

When we use  (\ref{eq:rt}), differently we keep the  pairs such that there is at least one exceedance and  the number of pairs  in $(\ref{CLvrais})$  is a random variable.
Moreover  we  refer to  conditional events.
The associated density is given by 
\begin{eqnarray*}
f_{s_i,s_j}(y_{ti},y_{tj};\theta)&=&- \frac{1}{\log G_{s_i,s_j}(u,u;\theta)}\times\\
&&
\left[\frac{\nabla_{ab}G_{s_i,s_j}(y_{ti},y_{tj};\theta)-
\nabla_{a}G_{s_i,s_j}(y_{ti},y_{tj};\theta)\nabla_{b}G_{s_i,s_j}(y_{ti},y_{tj};\theta)}{G_{s_i,s_j}(y_{ti},y_{tj};\theta)^2}\right]
\end{eqnarray*}
for $(y_{ti},y_{tj})\notin (-\infty,u]^2$.
In this case the components of the parameter vector $\theta$ are the marginal
parameter $\sigma_u$ and the dependence parameters $\beta$. 

Exceedances above $u$ will appear in
spatial clusters and in this respect the weights $w_{ij}$ in (\ref{CLvrais}) should  handle pairwise dependencies
inside and outside of the clusters.

Composite log-likelihood, as linear combination of log-likelihoods,
allows to obtain unbiased estimating equations under classical regularity conditions \citep{Heyde:1997}.
So, in principle, 
 it would be possible to come up with an optimal way of combining the individual scores $\nabla_\theta 
\log f_{s_i,s_j}(y_{ti},y_{tj};\theta)
 $ and determining  optimal weights $w_{ij}$ \citep{Kuk:2007}.
In the next section , more simply,    we look at weights for  reducing the  computational  burden as in \citet{Padoan:Ribatet:Sisson:2010}, like as the cut-off weights, namely   $w_{ij}=1$ if $\|s_i-s_j\|\le \delta$,
  and $0$ otherwise, for a fixed $\delta > 0$.
  There is some evidence for  Gaussian random processes   \citep{Davis:Yau:2011,Bevilacqua:Gaetan:Mateu:Porcu:2012}  that this choice 
  improves not only the computational efficiency but also the statistical efficiency.

{Under suitable conditions \citep[see the Appendix in][]{Padoan:Ribatet:Sisson:2010} the maximum composite likelihood estimator for $\theta$
can be proved consistent and asymptotically Gaussian, for large $T$.
 The asymptotic variance is given by the inverse of the Godambe information matrix
\begin{equation}\label{eq:godambe}
 \mathcal{G}_T(\theta)=\mathcal{H}_T(\theta)[\mathcal{J}_T(\theta)]^{-1}\mathcal{H}_T(\theta), 
\end{equation}
where $\mathcal{H}_T(\theta)=\E(- \nabla^2 cl_T(\theta))$ and $\mathcal{J}_T(\theta)=\Var(\nabla cl_T(\theta))$.
}

{ 
Standard error evaluation  requires consistent estimation of the matrices
$\mathcal{H}_T$ and $\mathcal{J}_T$.
Assuming strong-stationarity in time, we get 
 $$\mathcal{H}_T(\theta)=-T\E\left\{\nabla^2 U_1(\theta)\right\},$$
that can be  estimated by 
${{H}}_T=-\nabla^2 cl_T(\widehat {\theta}_T)$.
Estimation of the  matrix
$$\mathcal{J}_T(\theta)=\sum_{t=1}^T\sum^T_{k >t}\Cov\left\{\nabla U_t(\theta)\nabla U_{k}(\theta)^\prime\right\}$$
 requires some care.
   When we deal with real data, for instance environmental data, exceedances are seldom  independent in time. In such case declustering
techniques \citep{Nadarajah:2001} could be used to overcome the dependence in
time. An alternative way is 
employing a 
subsampling technique \citep{Carlstein:1986} to estimate  $\mathcal{J}_T$.
The subsampling method consists of  estimating  $\mathcal{J}_T$  
over $M$  overlapping temporal windows $D_j\subset\{1,\ldots,T\}$, $j=1,\ldots,M$,  of size $d_j$ by using the expression
$$
{{J}}_T=
\frac{T}{M}\sum_{j=1}^M \frac{\nabla cl_{D_j}(\widehat{\theta}_T)\nabla cl_{D_j}(\widehat{\theta}_T)^\prime}{d_j},
$$
 where  $\nabla cl_{D_j}$ is the composite likelihood score evaluated over the window $D_j$. An estimate of the asymptotic variance is then given by
$$
V_T=H_T^{-1}J_T H_T^{-1}. 
$$
}

\section{Numerical results}
\label{sec:simulation}
In this section we describe a simulation study with the aim of examining
and comparing the performances of CL estimates for the parameters of spatial max-stable processes.  Three versions of the CL
 are contrasted: 
two based on threshold exceedances
 that use the bivariate distributions (\ref{eq:lt}) and (\ref{eq:rt}), noted LT and RT, respectively, and 
one based  on block maxima data \citep{Padoan:Ribatet:Sisson:2010}, PRS hereafter. 
  The advantage of threshold methods 
  over the block maxima one is well known 
even if this advantage deserves to be evaluated \citep{Madsen:Rasmussen:Rosbjerg:1997}.

 We have considered the Gaussian extreme value process 
 (\ref{eqSmith})
 and we have set 
$\displaystyle{\Sigma=\left( \begin{array}{ll}
200 & 150\\
150 & 300
\end{array}
\right)}$
as in \citet{Padoan:Ribatet:Sisson:2010}. 
The spatial process is observed  on   $n^2$
points located over a $n\times n$ regular grid $\{k,\ldots,n*\,k\}^2$,
where $n=5,7$. Here we  present only the results for the 49 sites, because 
evidences are virtually the same with the smallest grid { (25 sites)}.
Three lags  $k$ for the spatial sites, namely  $k= 1, 5, 10$, are considered in order to take into account different levels
of the spatial extremal dependence. 

The spatial extremal dependence of the maxima can be  characterized using the extremal coefficient function \citep{Smith:1990, Schlather:Tawn:2003}, $v(\cdot)$, given by
$$
\Pr\left(\max(Z(s),Z(s+h))\le z\right)=
\exp\{-v(h)/z\}.
$$
For any vector $h$, we have $1\leq v(h)\leq 
2$ and $v(h)=1$ corresponds to the perfect dependence, instead for independent maxima we have $v(h)=2$. In the case of the  Gaussian 
extreme value process, the  extremal coefficient function is 
 $$v(h)=2\Phi\left(\frac{\sqrt{h'\Sigma^{-1}h}}{2}\right). $$
 Figure \ref{fig:ens49nt10-25} 
 shows the  values of this function for the grids over 49 spatial 
sites and for the different spatial lags. As expected, the spatial 
extremal dependence decreases as the spatial lag increases. For $k=1$, the extremal dependence keeps 
strong on the whole grid. For $k=10$, the dependence appears strong only  for
neighbouring sites.

 In our experiments, for each simulation,  we have considered $40$ years and  for each year we have a sample of $91$ independent daily observations.
Our aim is to mimic a framework where   maxima are taken over a fixed season of the year. 
Then we have considered $500$ simulations  for evaluating  mean,  standard deviation, and root mean square error of the estimates.

Simulations have been carried out 
using the \texttt{rmaxstab} function of the R package \texttt{SpatialExtremes} \citep{SpatialExtremes}.
This function allows to  simulate a finite realization $(Z(s_1),\ldots,Z(s_p))'$ of a Gaussian extreme value process  with unit 
Fr\'echet marginal distributions.
In order to fulfill the condition 
   $0<G_{s_i,s_j}({0},0)<1$ 
of Theorem 2.1 in  \citet{Rootzen:Tajvidi:2006},
we have  transformed the data to the Gumbel scale, namely  $Z^*(s_i)=\log(Z(s_i))$, 
$i=1,\ldots,n$.  The threshold $u$ for all sites has been set equal to the empirical $0.98$-quantile, calculated over all data in each Monte Carlo simulation.

As mentioned at the end of Section \ref{sec:inference}, a careful choice of the weights in  (\ref{CLvrais})
could improve the statistical efficiency in addition to the computational one. 
In fitting the models, we have  considered three sets of weights, according to  different values of $\delta$,
where  $\delta$ has been set equal to the 
$a$-quantile, $q_a$,    of the distances among all pairs of sites and $a=0.25, 0.50, 1.00$.

{
Figures (\ref{fig:comparison-sigma11}), (\ref{fig:comparison-sigma22}) and
(\ref{fig:comparison-sigma12})
 contrast the variability of the estimates  for grids with different  spatial lags and  different set of weights derived from  
 $\delta=q_{0.25}$ and  $\delta=q_{1.00}$. {The RT estimates  always display the smallest variability but raise some concerns according different degrees of  spatial dependence. When we use 
all the pairs ($\delta=q_{1.00}$) in forming the CL function,
the  bias  of the $\sigma_{11}$ and $\sigma_{22}$ estimates seriously increases in case of  weak spatial dependency (spatial lag $k=10$). On the other hand  LT results point out  an efficiency  similar to the PRS method  and the bias of the estimates keep within the reasonable bounds, regardless of the spatial dependence   and the number of the pairs. 
 }

{In 
Tables  \ref{tab:49-1}, \ref{tab:49-5}  and \ref{tab:49-10} we  measure the efficiency of  the estimates in terms of  the root mean square errors  (RMSEs)
under  different extremal dependence and different amount of pairs.   Bold figures highlight  the minimum value of RMSEs.  
When the extremal dependence is strong  (grids with 
lag $k=1$), the RT method clearly outperforms the LT one for whichever value of  $\delta$. For moderate spatial 
dependencies (grids with lags $k=5$), there is again an advantage in prefering the RT method with respect to
the LT. 
 Results for weak spatial 
 dependencies (grid with lag $k=10$) are quite differents and put forwards the role of the number of pairs: for 
 $\delta=q_{0.25}$, RT approach gives smallest RMSE than LT and for  $\delta=q_{1.00}$, the converse happens. For 
  $\delta=q_{0.50}$, there is not clear indication of which method behaves better.} 

{However our results, consistent to the  literature on weighted composite likelihood
 \citep{Joe:Lee:2009,Davis:Yau:2011,Bevilacqua:Gaetan:Mateu:Porcu:2012},  
display that 
including many pairs in the pairwise likelihood can harm the CL estimator, suggesting that we should retain as few pairs as possible.
}

{
We have already remarked that the RT  estimates always had   smaller standard errors  than the LT ones.
Nevertheless, they also
lead  to a larger RMSEs  when the spatial dependence is weak   and we use all the pairs because  a large bias is present. 
This observed bias in  the RT estimates seems also depend on the choice of the
threshold. As matter of proof,  we did the same Monte Carlo experiment  setting the threshold $u$ equal to  the empirical  $0.90$, $0.95$ and $0.98$ quantiles of the data. 
These values correspond to the usual choice in the real data analysis.
Table \ref{tab:quantiles-distance} reports the bias  of the parameter estimates according the three  values of the threshold. 
We can see  that for lower  thresholds,  the absolute value of the bias increases for the 
$\sigma_{11}$ and $\sigma_{22}$ estimates following the RT approach. On the other hand 
the bias of the LT estimates is always moderate.}


\section{Discussion}
\label{sec:discussion}

{In this paper, we have compared two threshold exceedances models to estimate the parameters 
of a particular instance of spatial max-stable processes, namely the Gaussian extreme value process \citep{Smith:1990,Padoan:Ribatet:Sisson:2010}
of spatial 
max-stable processes.
Each model is based on a particular bivariate extreme value property  \citep{Rootzen:Tajvidi:2006,  Ledford:Tawn:1996}.  
However our estimation strategy can be easily extended to other spatial max-stable processes \citep
{Schlather:2002,Kabluchko:Schlather:de_Haan:2009}, provided that the  expression of the bivariate
distribution is easy to evaluate.
}


{Our simulation experiments point out that 
both methods are valuable and
choosing between the RT and LT approach relies on the degree of the spatial 
dependence.
 When the spatial dependence between high values is strong, the RT 
approach seems preferable.
 The RT approach
for weak spatial dependent data
displays  a significative bias in the 
estimation. 
On the other hand LT approach does not suffer from this problem.
}

{A possible explanation of this evidence is that  the bivariate distribution (\ref{eq:rt}) deals with  pairs of observations with at least one component  exceeding the  threshold $u$. As \citet{Rootzen:Tajvidi:2006} have underlined in their paper, the multivariate generalized Pareto is degenerate in the 
case of independence, so it is not surprising that the weak extremal dependence can spoil the CL estimates.
In this regard, our findings are consistent with the results obtained  by \cite{Rakonczai:Tajvidi:2010} for a bivariate logistic extreme value  with weak dependence. Moreover, the 
distribution (\ref{eq:rt})  is not the true conditional bivariate distribution but only its asymptotic approximation.  Again in the case of weak extremal dependence, the convergence may be slow, leading  to worry regarding to the usual behaviour of the CL estimates. 
}

{Nevertheless our simulations  highlight that a clever choice for the weights of the CL could significantly improve the efficiency of the 
estimates. 
This is especially true under weak spatial dependency, where the RT approach gives the best RMSE results.
Indeed, when we consider all the pairs, pairs of sites which are not 
spatially dependent could be linked because of an exceedance of one  component. In the bivariate CL framework, this should lead to a misleading link in  the extremal association and, as a consequence, contribute to an incorrect { estimate}.} 

{This fact entails that there is a effective need of simple practical rules for fixing the weights.
However rules based only on the  extremal 
coefficient function appear as not really adapted and more theoretical work seems necessary.
 A possible solution    that we will explore in the future
 is   to choose the  weights
 by minimizing a certain norm of the Godambe information 
 matrix   (\ref{eq:godambe}) as in \citet{Bevilacqua:Gaetan:Mateu:Porcu:2012}.}

\bibliography{ext}
\bibliographystyle{ECA_jasa}

\newpage

\begin{figure}[th!]	
\begin{tabular}{cc} 
&$k=1$  \\
 \includegraphics[width=7.0cm,height=6cm]{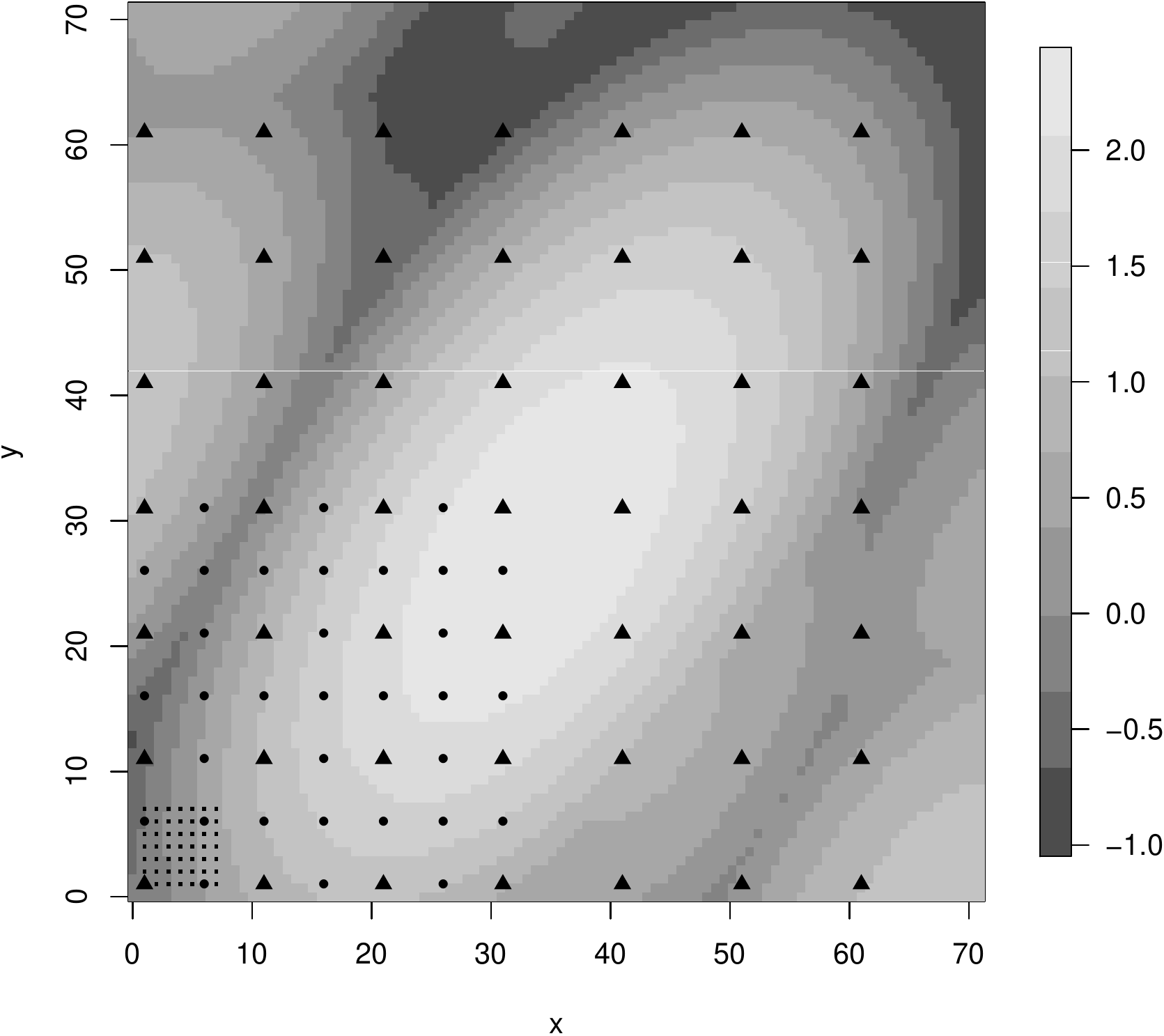}
 &
 \includegraphics[width=6cm,height=6cm]{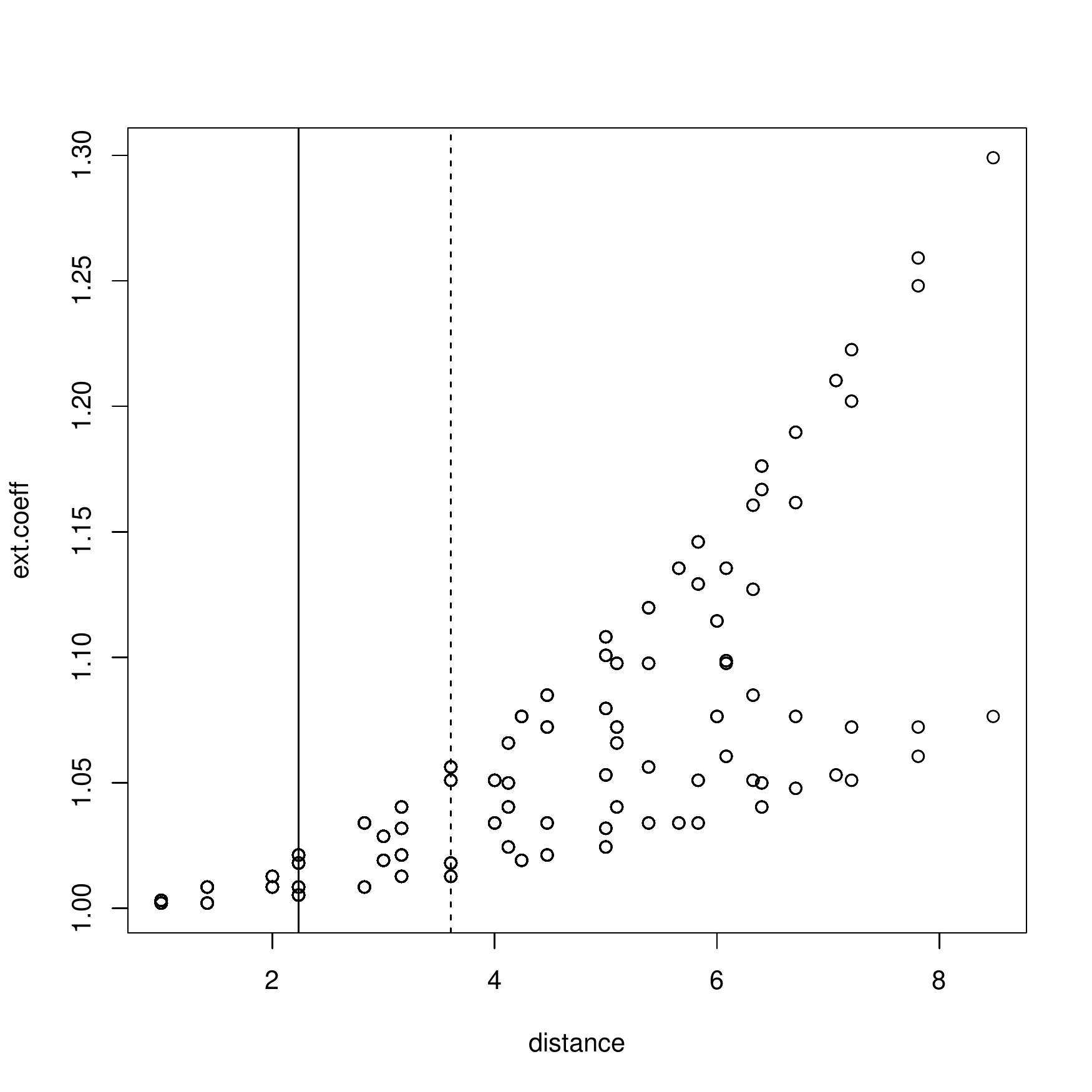} \\
 $k=5$ & $k=10$ \\
\includegraphics[width=6cm,height=6cm]{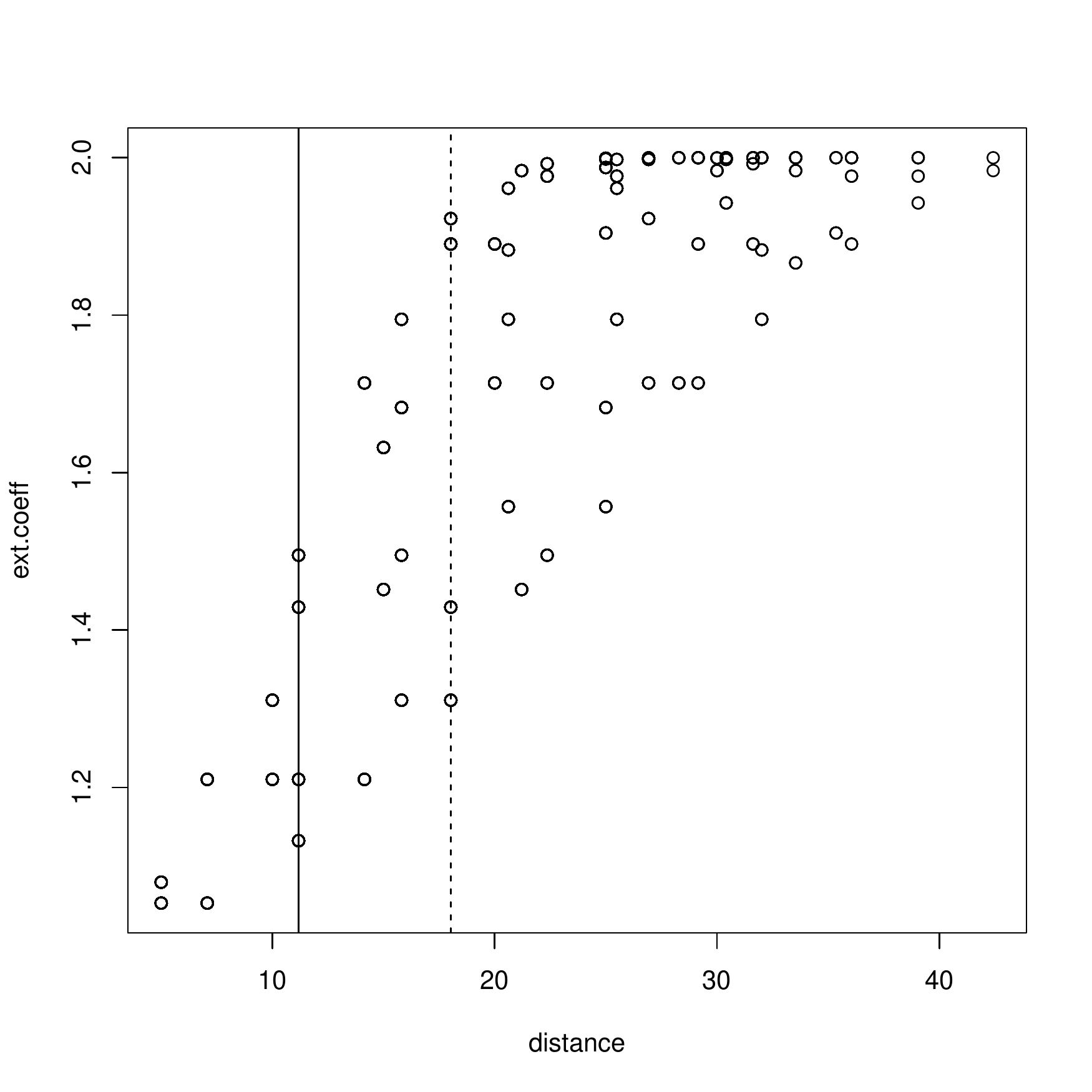} &
\includegraphics[width=6cm,height=6cm]{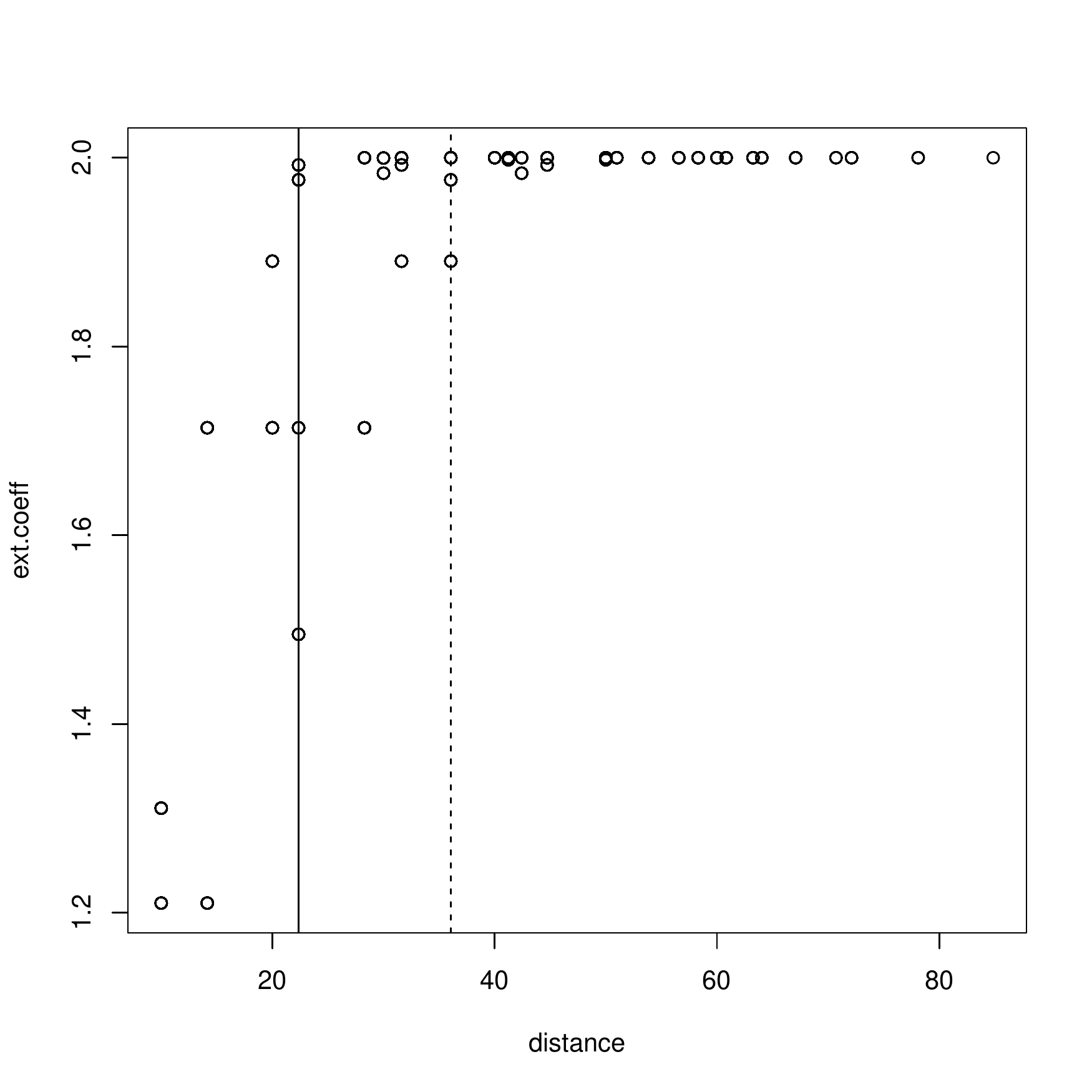}
\end{tabular}
\caption{On the left top corner: an example of a simulation from a Gaussian  extreme value process, with 
$\sigma_{11}=200$, $\sigma_{22}=300$, $\sigma_{12}=300$.
Three different regular grids, with different  spatial lags  ($k=1,5,10$), are superimposed.
The other figures are the plots of the extremal coefficient function $v(h)$. This function is evaluated at the  distances among  49 sites on  regular grids with different  spatial lags.
Solid lines (dashed lines) correspond to the value $\delta=q_{0.25}$ ($\delta=q_{0.50}$, respectively).}
\label{fig:ens49nt10-25}
\end{figure}

\begin{figure}[h!]
\begin{tabular}{cc}
$\delta=q_{0.25}$ &$\delta=q_{1.00}$ \\
\includegraphics[width=6cm,height=6cm]{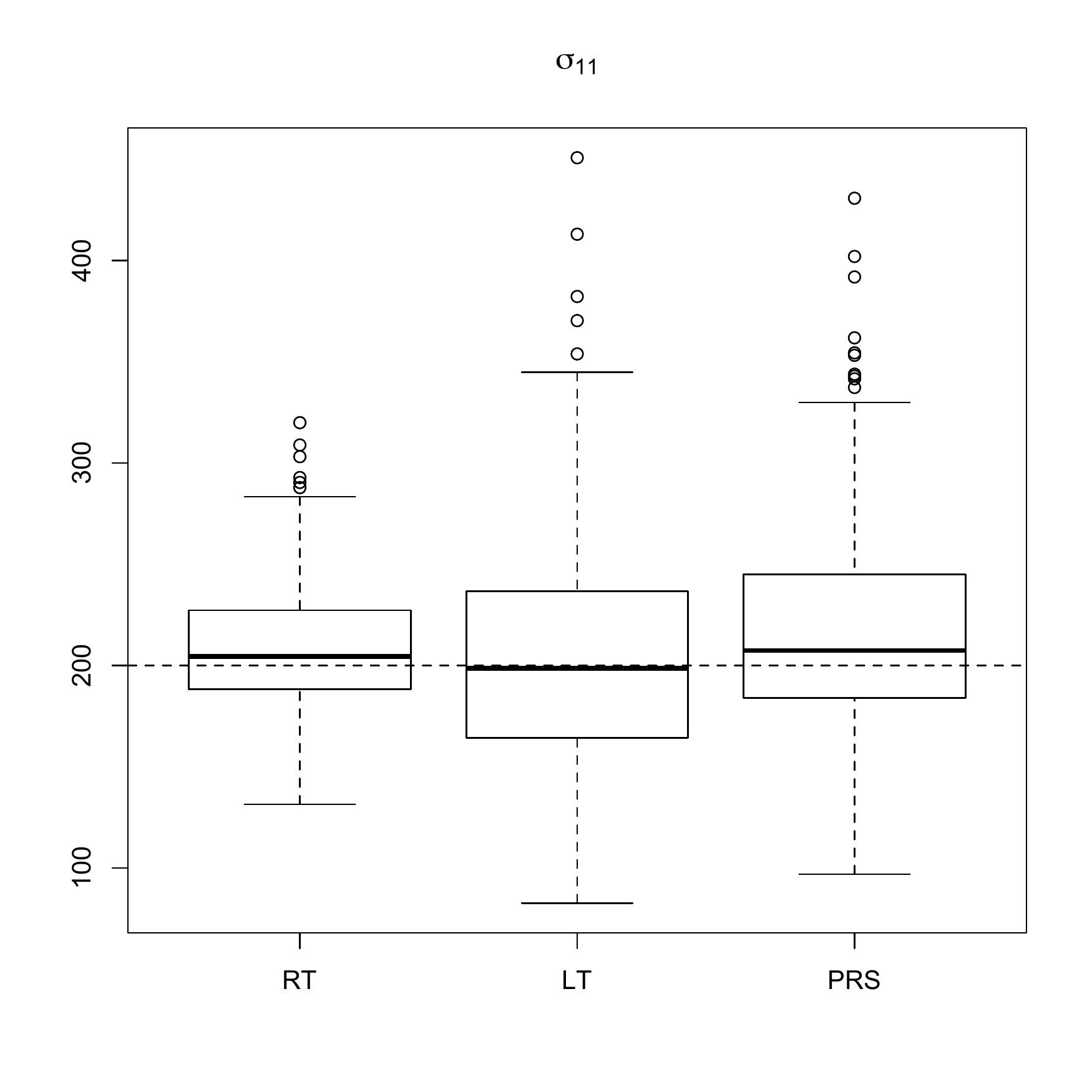} 
&
\includegraphics[width=6cm,height=6cm]{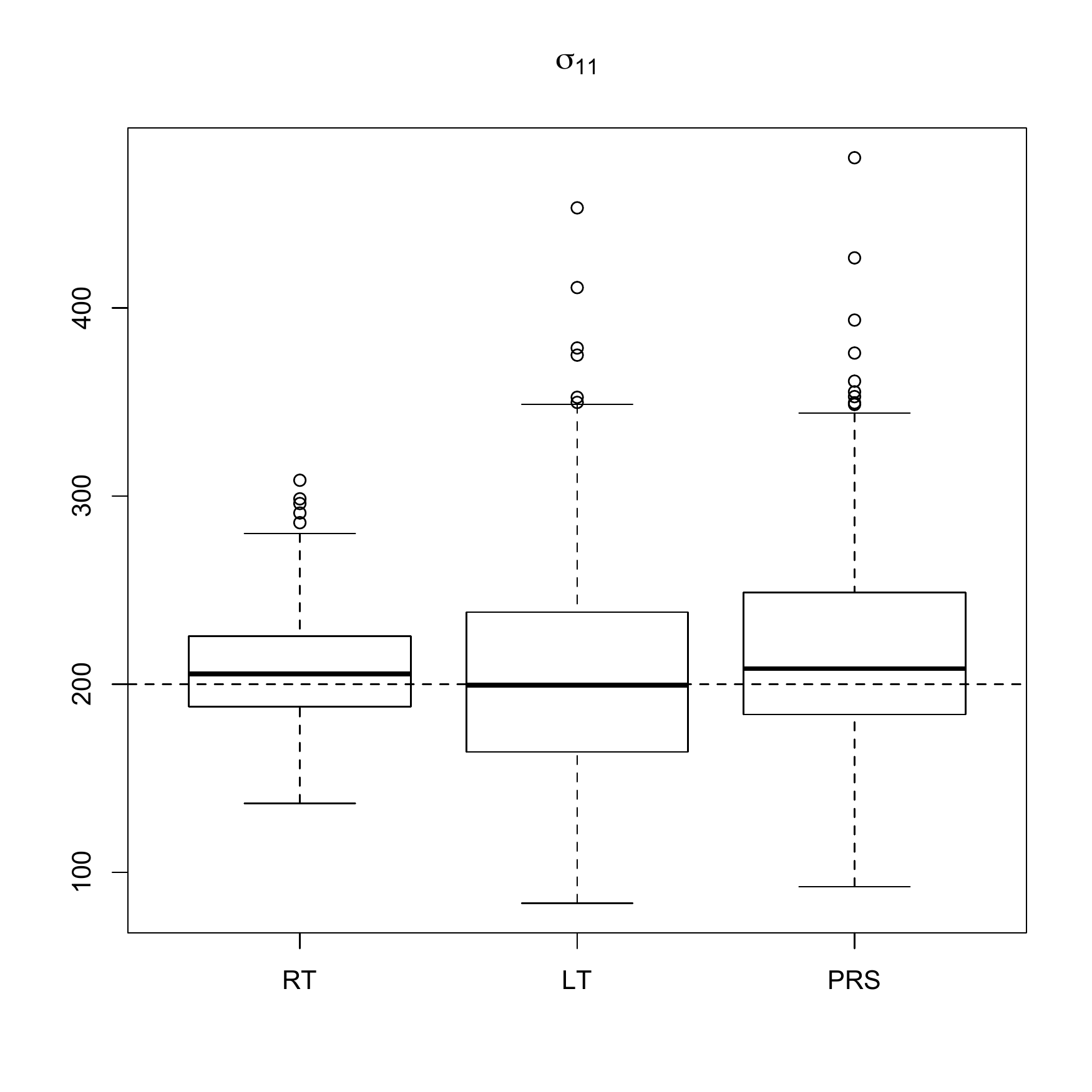} 
\\[-0.2cm]
\includegraphics[width=6cm,height=6cm]{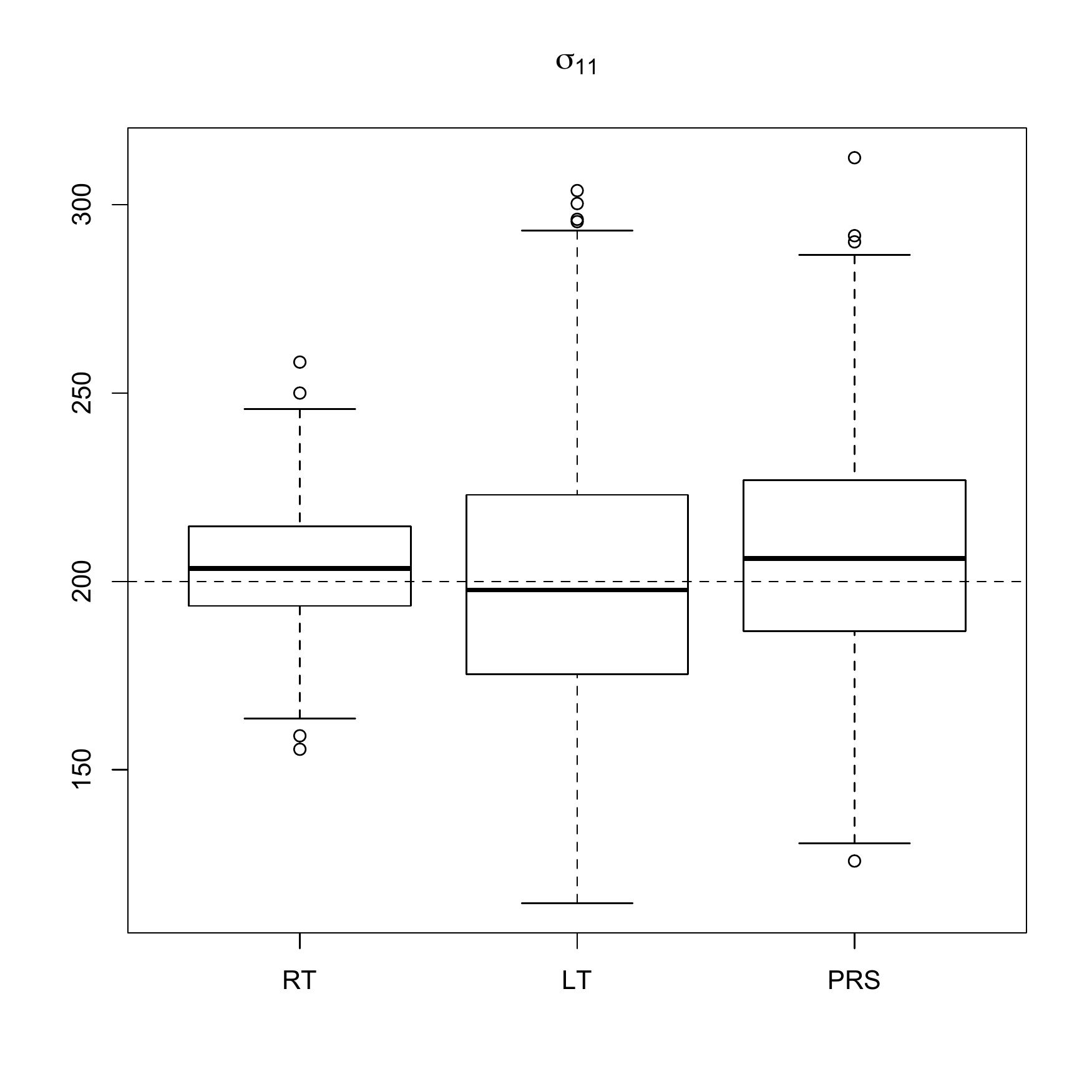} 
&
\includegraphics[width=6cm,height=6cm]{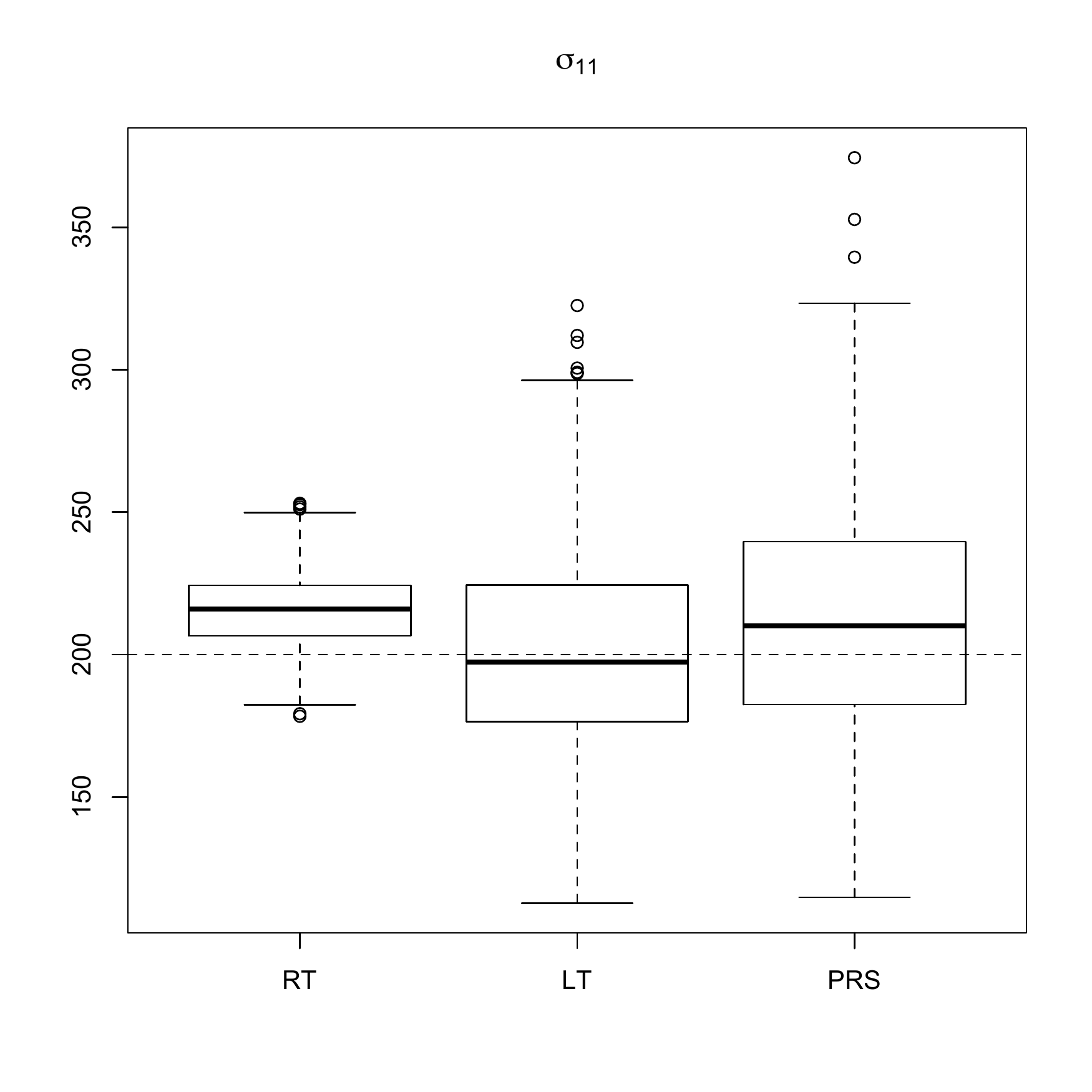} 
\\[-0.2cm]
\includegraphics[width=6cm,height=6cm]{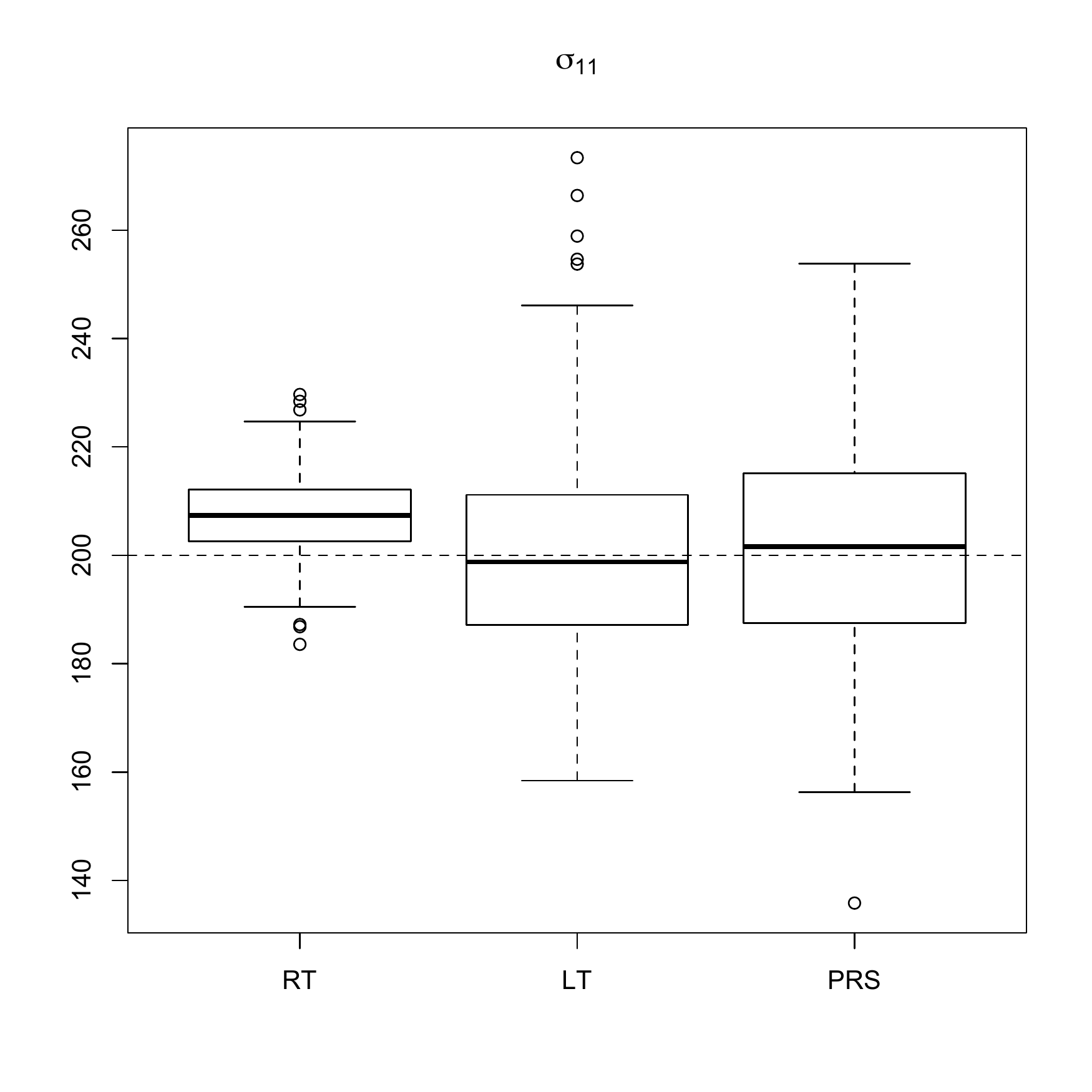} 
&
\includegraphics[width=6cm,height=6cm]{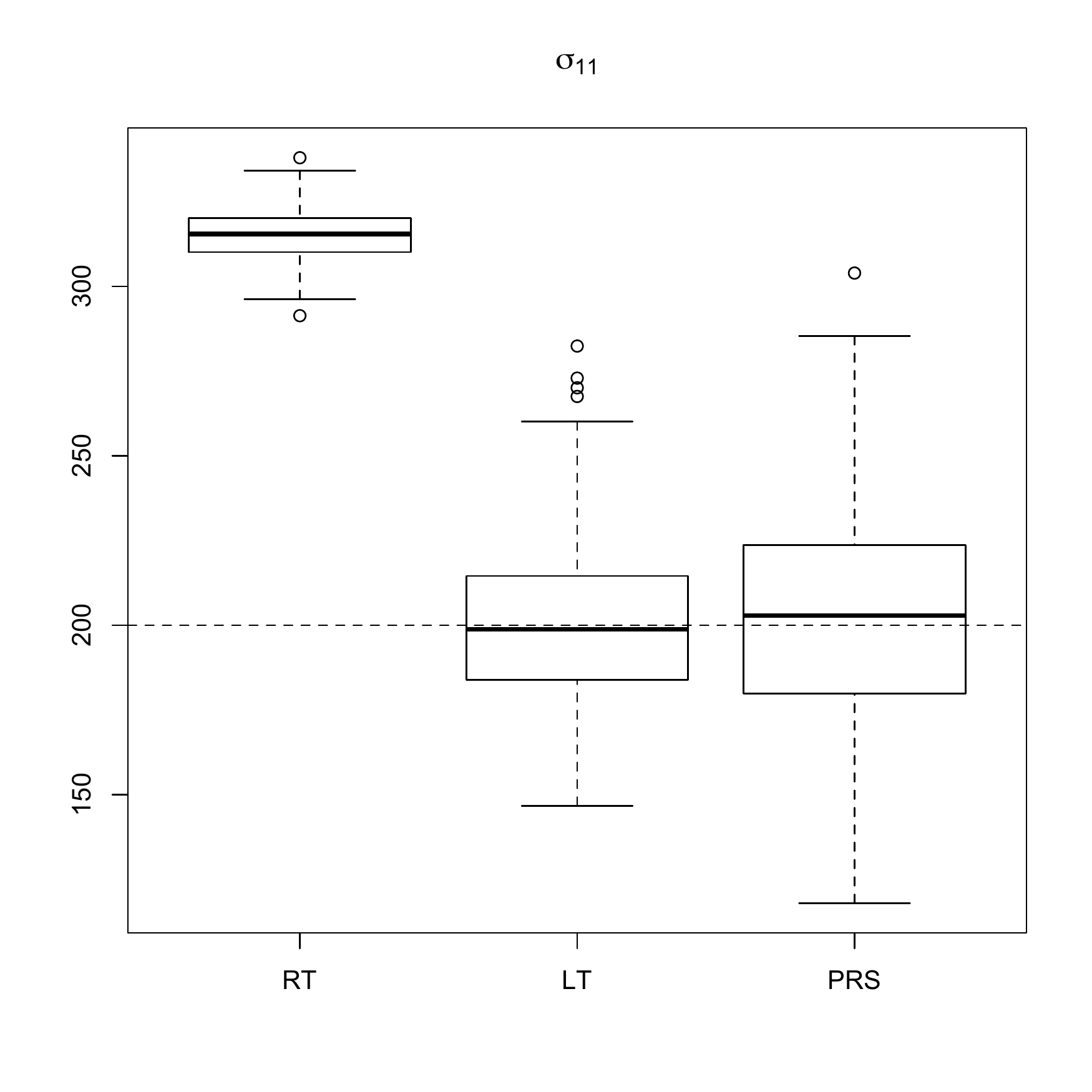} 
\end{tabular}
\caption{ Box-plots of the  estimates of $\sigma_{11}$.
The true value for $\sigma_{11}$ is indicated  by a dotted line. Each row in the panel indicates the results over a grid with
lag $k=1,5,10$,  respectively. 
In the first (resp. second) column of the panel the composite likelihood weights correspond to $\delta=q_{0.25}$ (resp. 
$\delta=q_{1.00}$).}
\label{fig:comparison-sigma11}
\end{figure}

\begin{figure}[h!]
\begin{tabular}{cc}
$\delta=q_{0.25}$ &$\delta=q_{1.00}$ \\
\includegraphics[width=6cm,height=6cm]{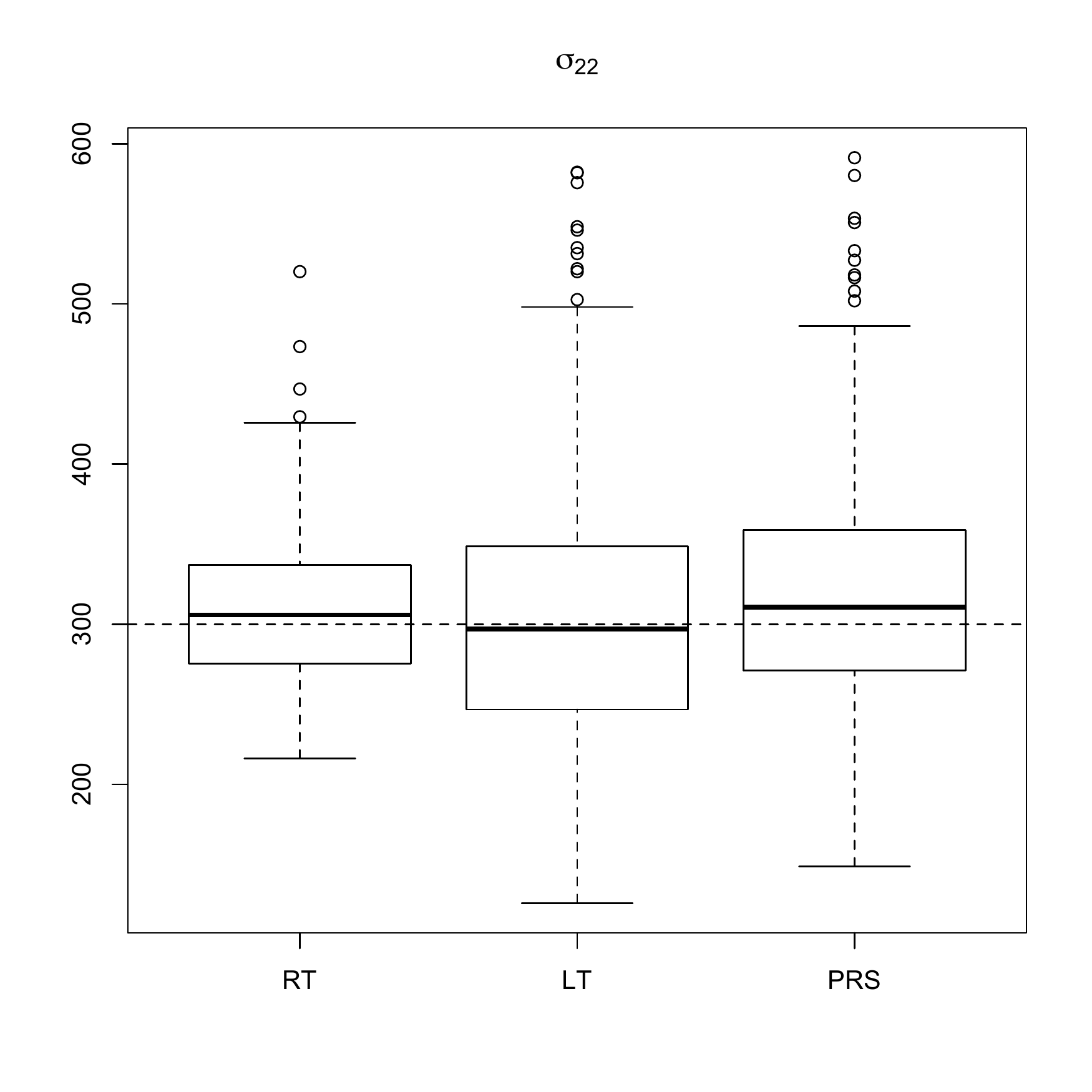} 
&
\includegraphics[width=6cm,height=6cm]{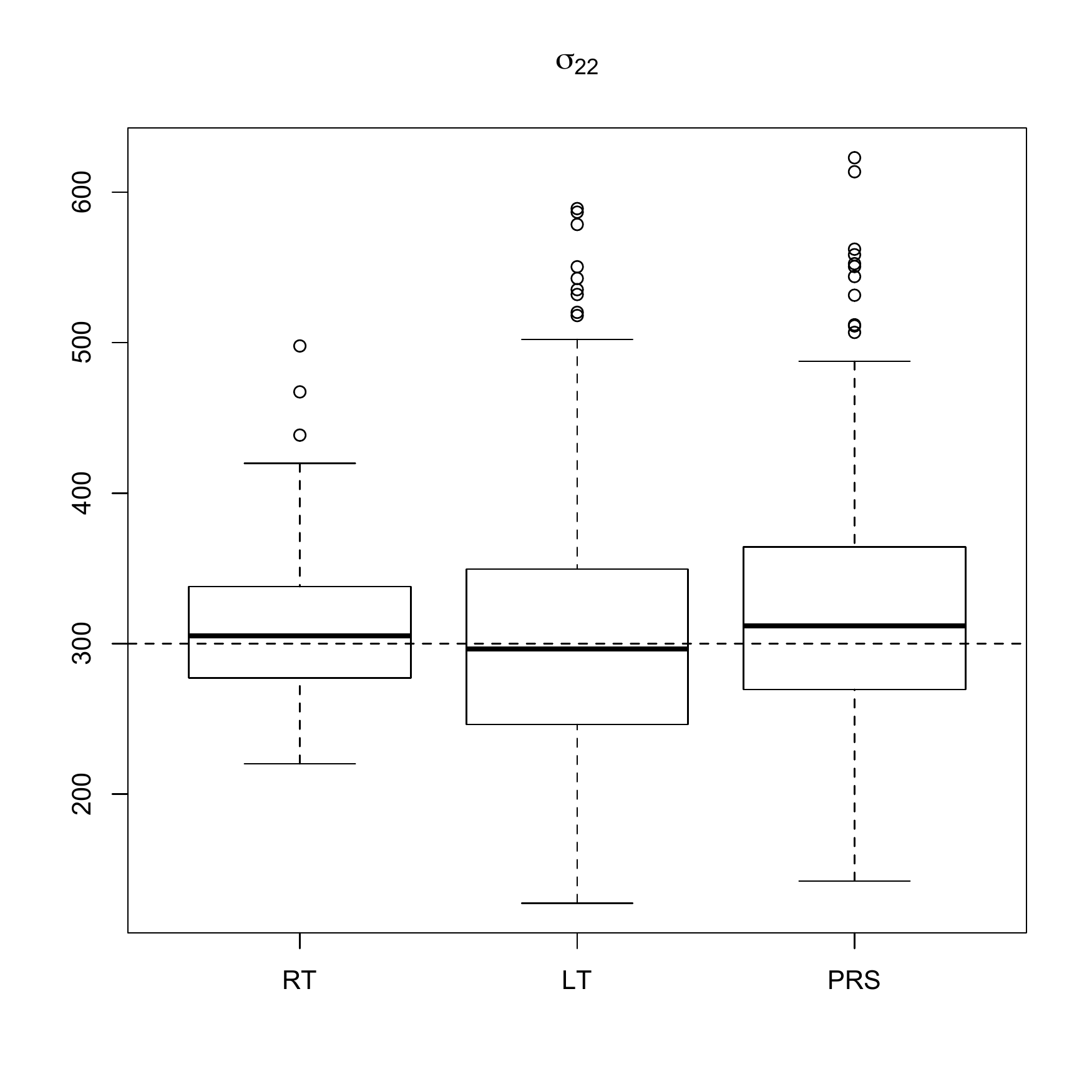} 
\\[-0.2cm]
\includegraphics[width=6cm,height=6cm]{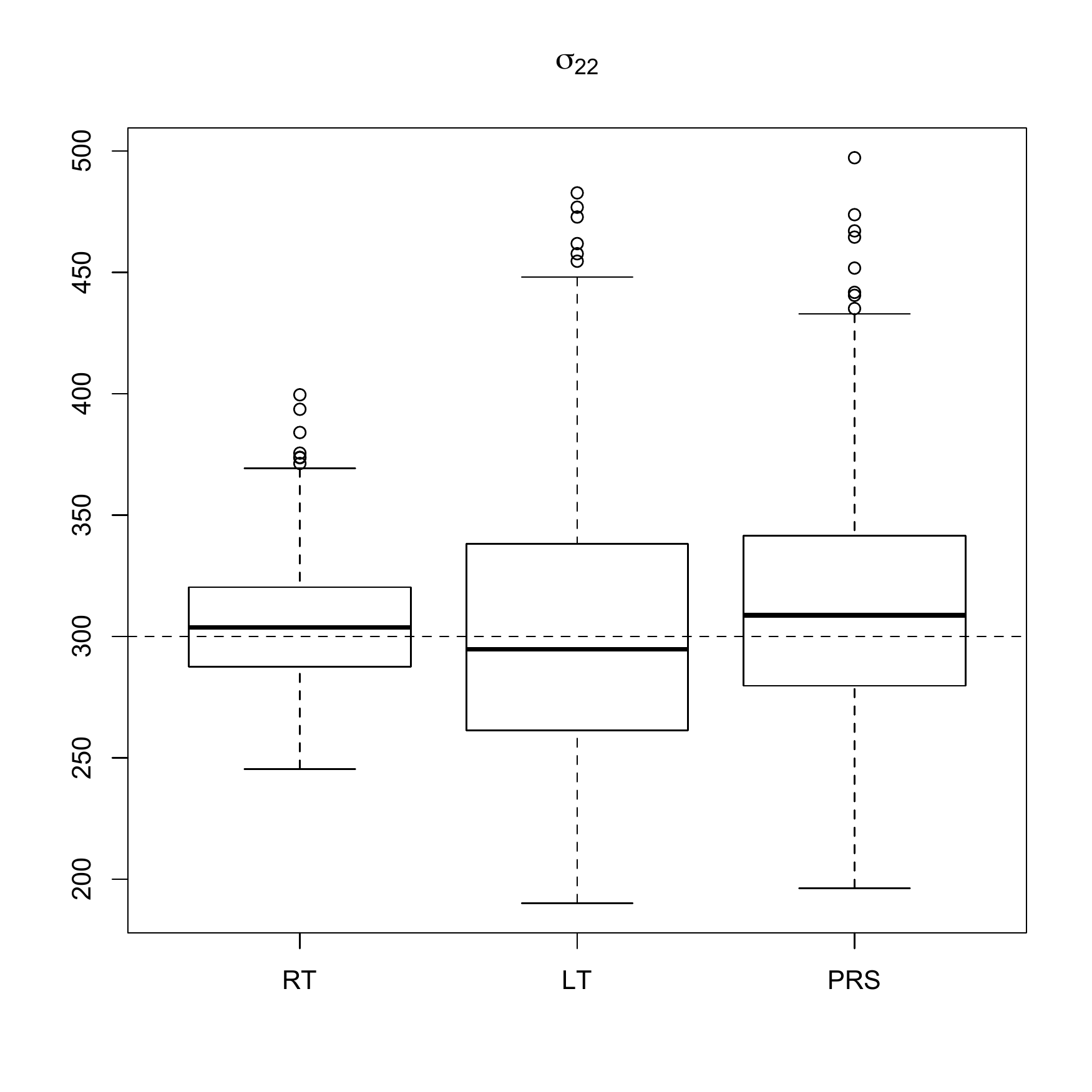} 
&
\includegraphics[width=6cm,height=6cm]{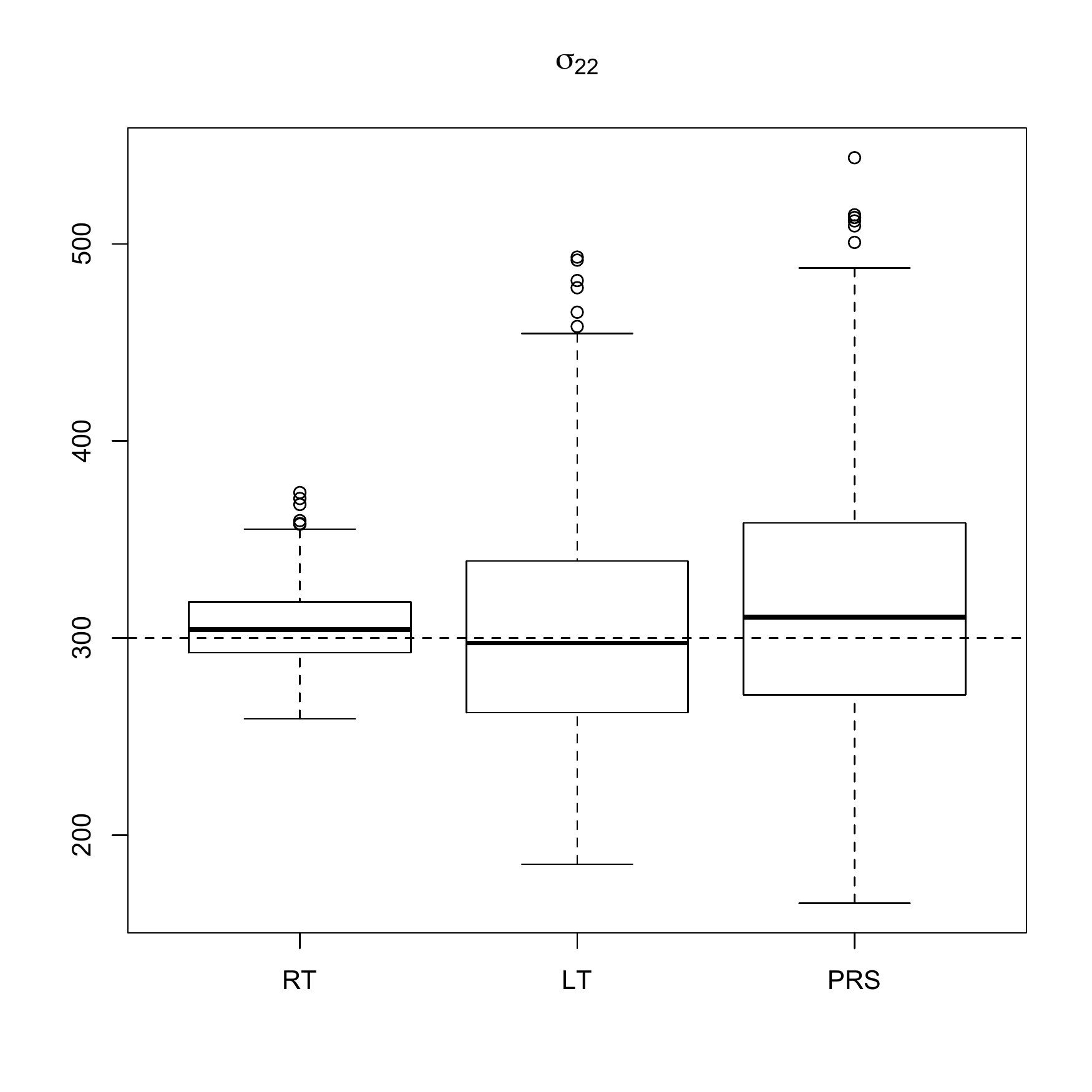} 
\\[-0.2cm]
\includegraphics[width=6cm,height=6cm]{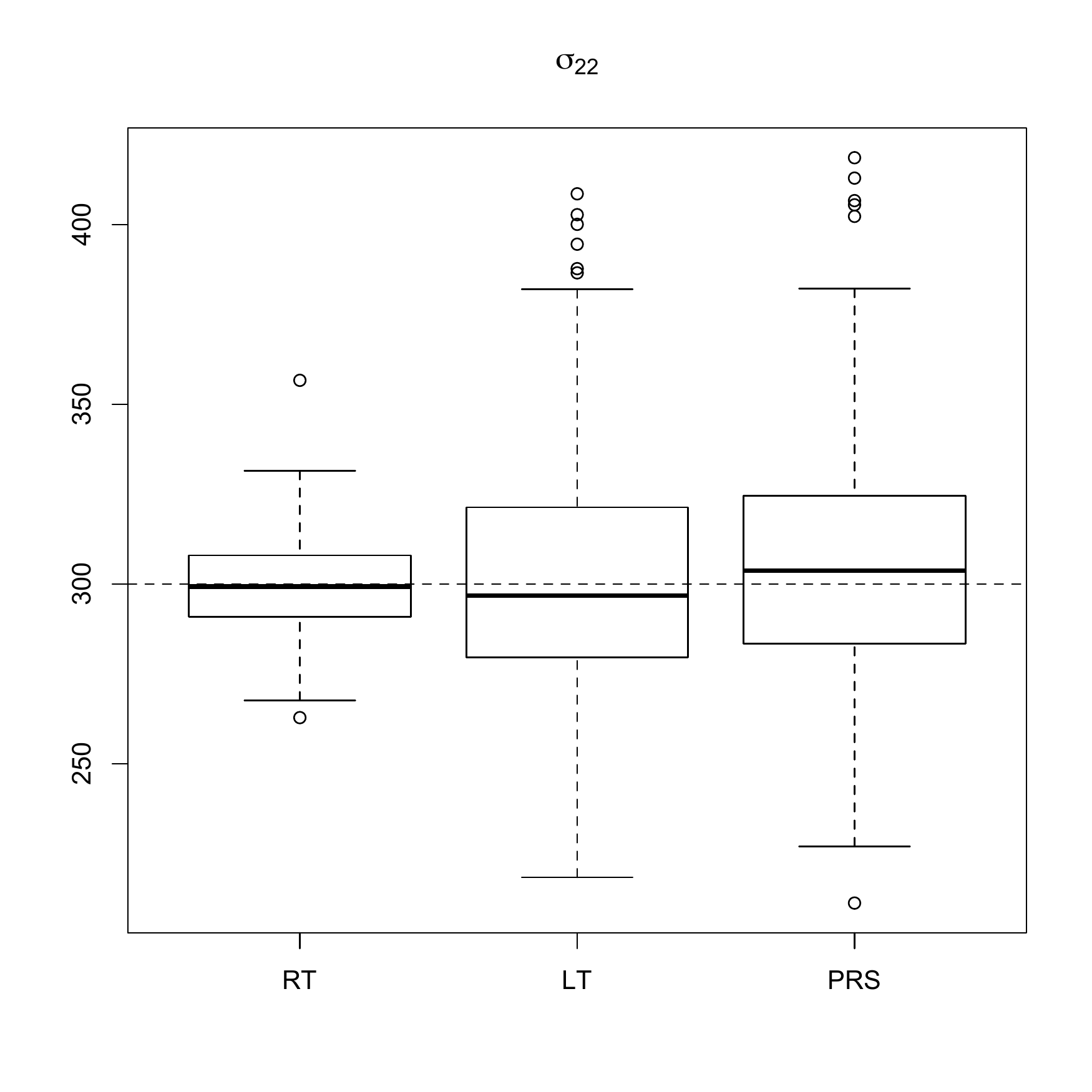} 
&
\includegraphics[width=6cm,height=6cm]{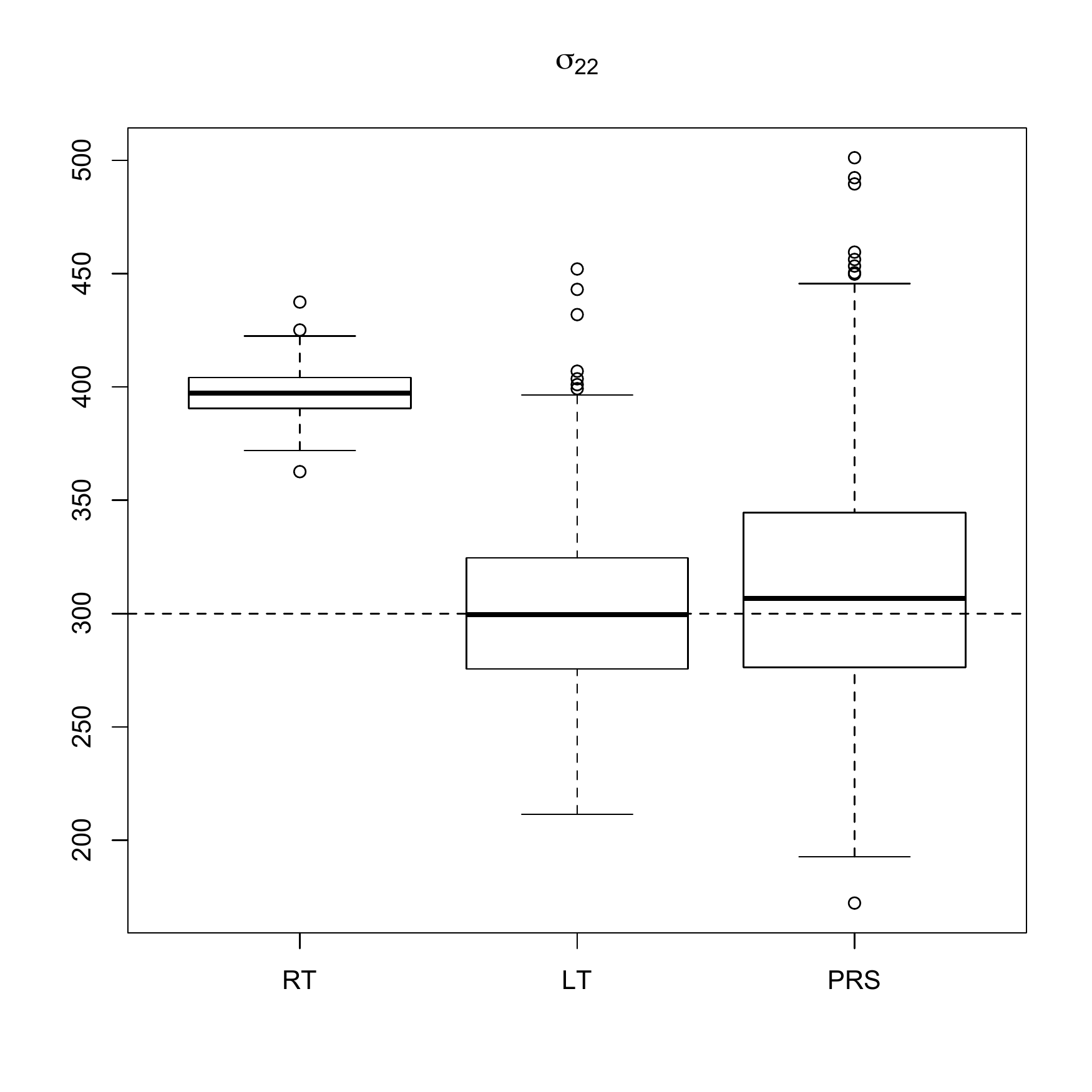} 
\end{tabular}
\caption{Box-plots of the estimates of  $\sigma_{22}$.
The true value for $\sigma_{22}$
is indicated  by a dotted line. Each row in the panel indicates the results over a grid with lag $k=1,5,10$, respectively. In the first (resp. second) column of the panel the composite likelihood weights correspond to $\delta=q_{0.25}$ (resp. 
$\delta=q_{1.00}$).}
\label{fig:comparison-sigma22}
\end{figure}

\begin{figure}[h!]
\begin{tabular}{cc}
$\delta=q_{0.25}$ &$\delta=q_{1.00}$ \\
\includegraphics[width=6cm,height=6cm]{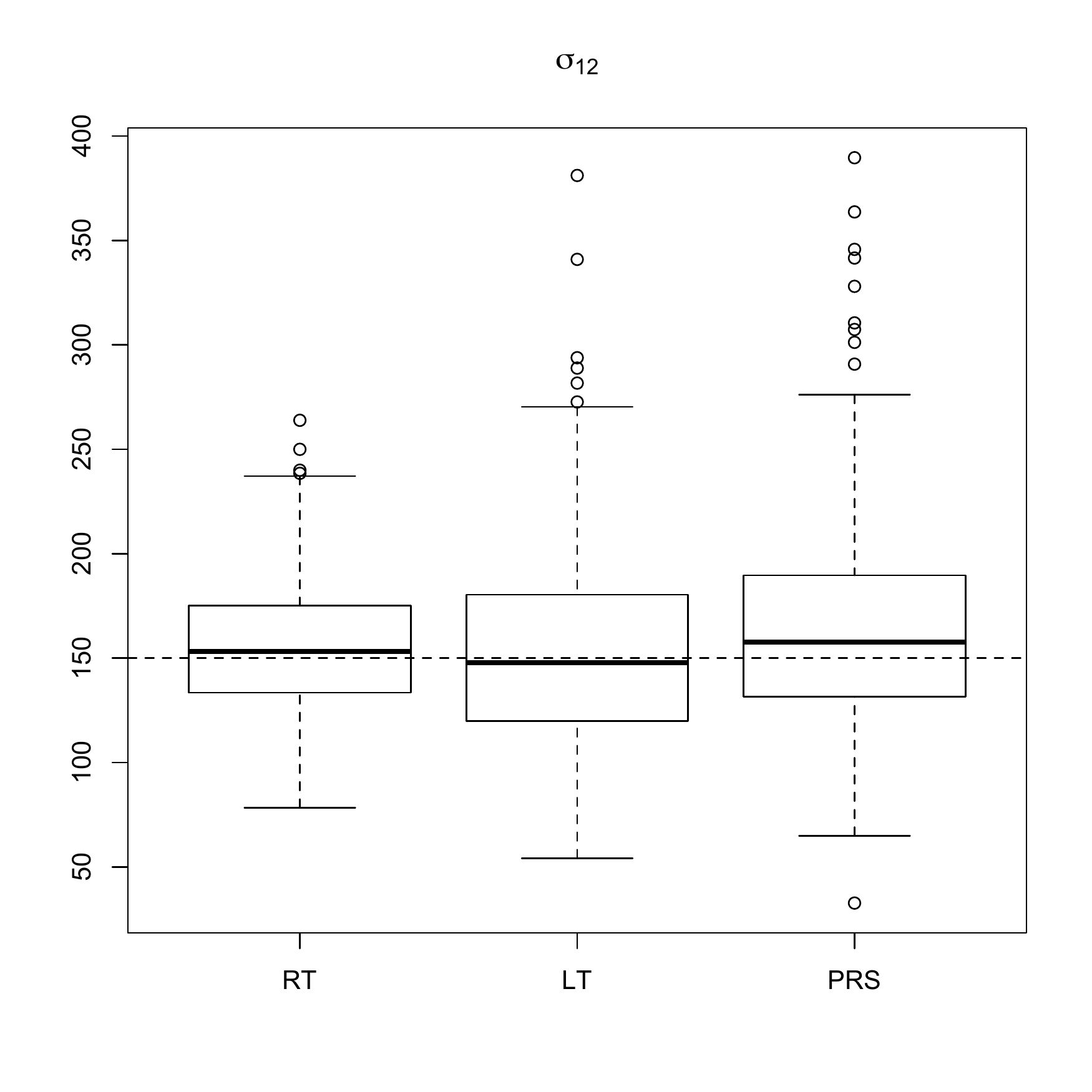} 
&
\includegraphics[width=6cm,height=6cm]{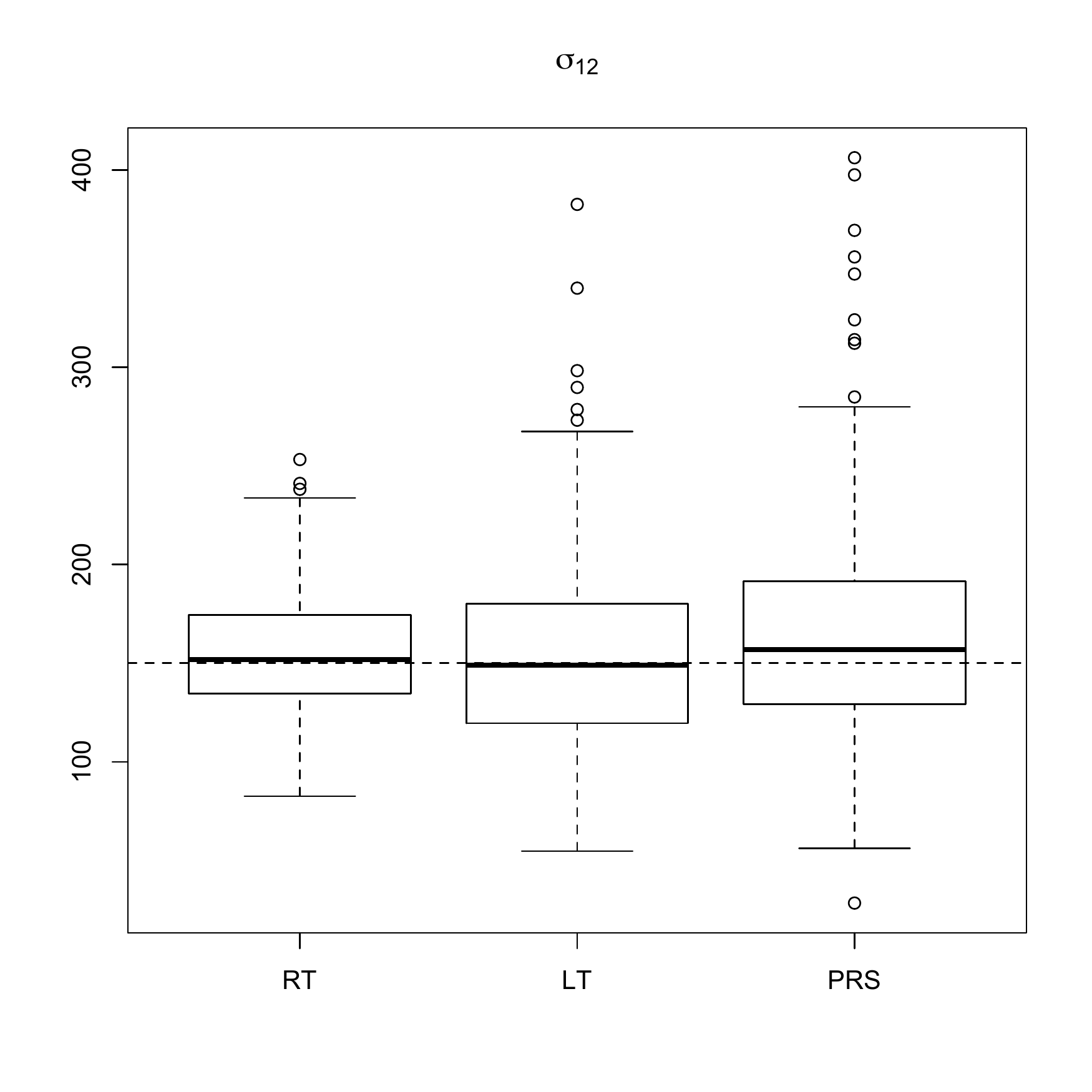} 
\\[-0.2cm]
\includegraphics[width=6cm,height=6cm]{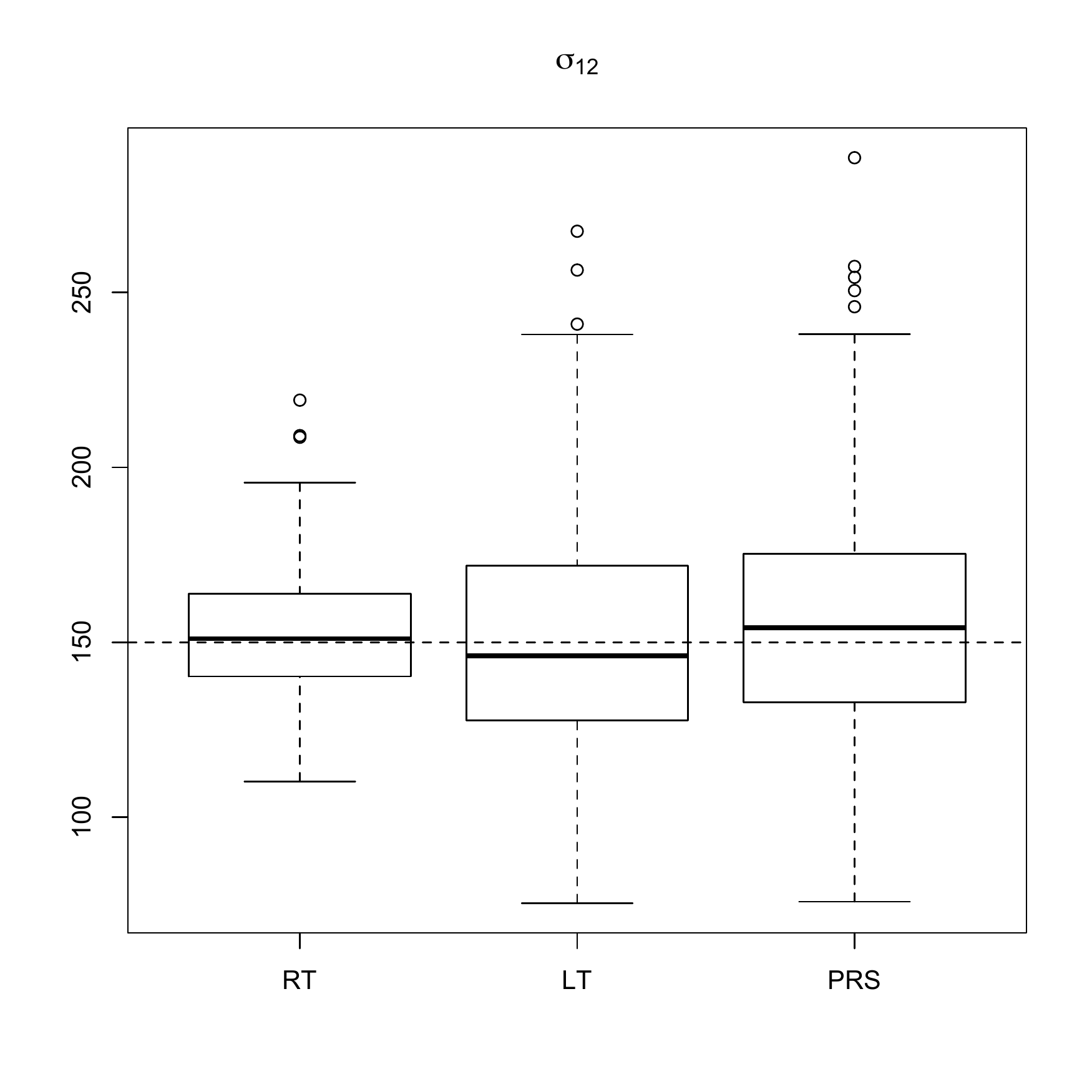} 
&
\includegraphics[width=6cm,height=6cm]{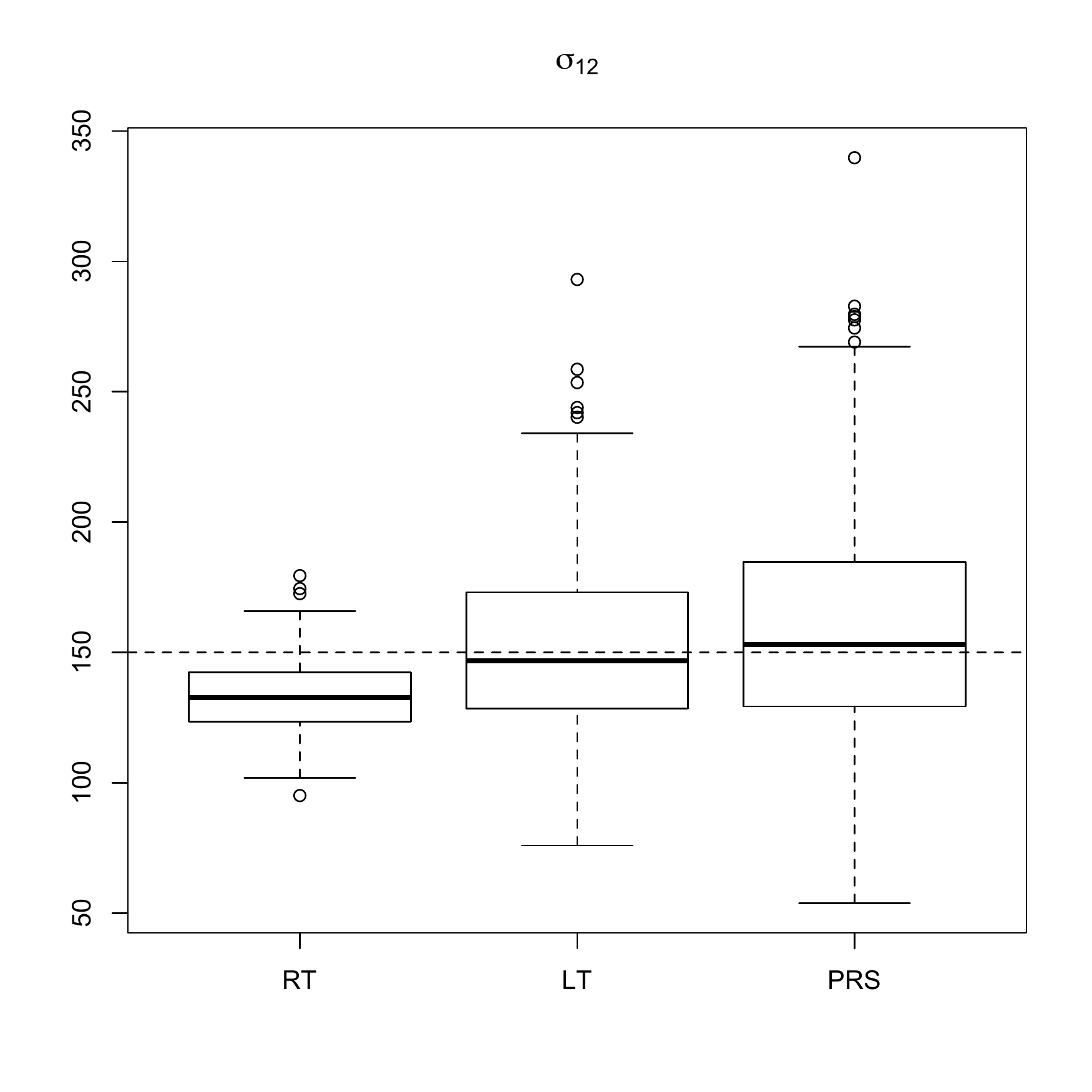} 
\\[-0.2cm]
\includegraphics[width=6cm,height=6cm]{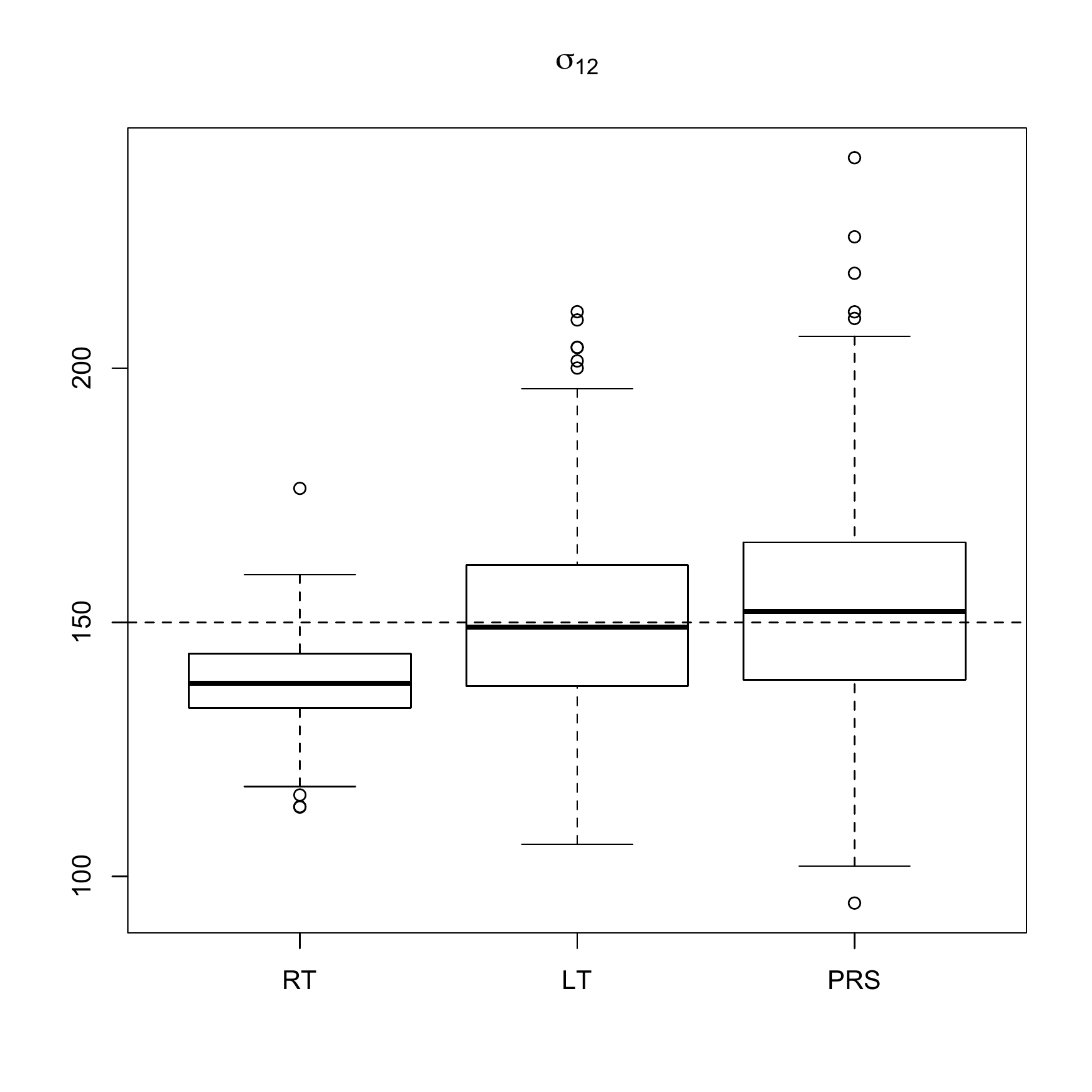} 
&
\includegraphics[width=6cm,height=6cm]{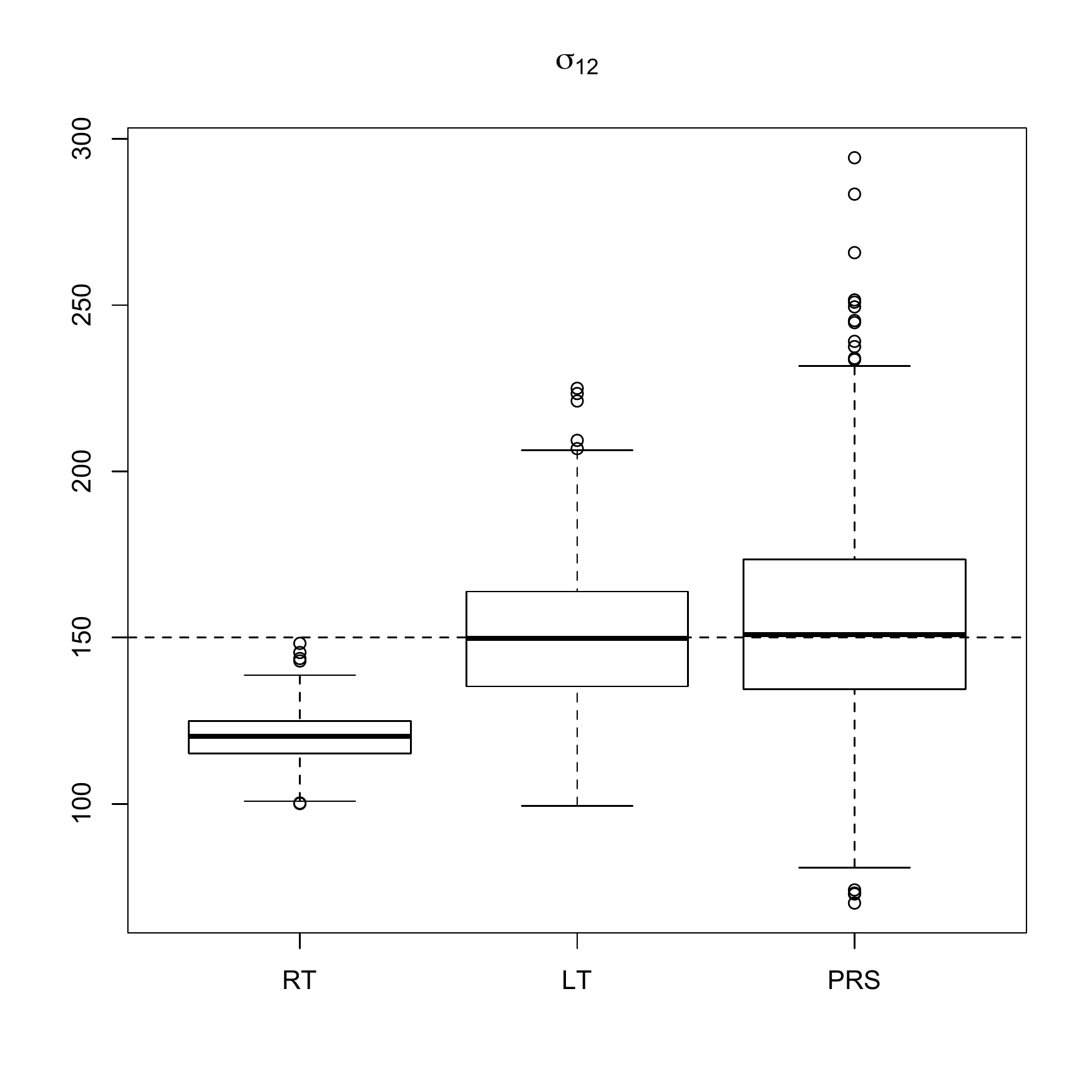} 
\end{tabular}
\caption{Box-plots of the estimates of $\sigma_{12}$.
The true value for $\sigma_{12}$
is indicated  by a dotted line. Each row in the panel indicates the results over a grid with lag $k=1,5,10$, respectively. In the first (resp. second) column of the panel the composite likelihood weights correspond to $\delta=q_{0.25}$ (resp. 
$\delta=q_{1.00}$).}
\label{fig:comparison-sigma12}
\end{figure}

\begin{table}
\begin{center}
\begin{tabular}{|r|r|r|r|r|}\hline
   & Parameter            & $\sigma_{11}$ & $\sigma_{22}$ & $\sigma_{12}$ \\ \hline
   &True value  & 200.00 & 300.00 & 150.00 \\\hline\hline
   $\delta=q_{0.25}$  & & & &\\\hline\hline
    &RT & 209.13 (29.62) &308.71 (44.14) & 156.49 (30.06)\\ 
 &  & {\bf 30.99} &  {\bf 44.99} &  {\bf 30.75} \\ \hline
  &LT & 204.24 (54.85) & 301.80 (80.63) & 152.71 (46.81)\\ 
  & &55.01 & 80.65 & 46.89 \\\hline
 & PRS & 216.94 (47.77) & 320.01 (67.19) & 162.03 (46.94)\\ 
 &  & 50.69 & 70.10 & 48.46\\ \hline\hline
 $\delta=q_{0.50}$  & & & &\\\hline\hline  
  &RT & 209.00 (29.18) & 308.46 (43.53) & 156.31 (29.57)\\ 
  &  &  {\bf 30.54} &   {\bf 44.34} &  {\bf 30.24} \\ \hline
     &LT & 204.75 (54.35) & 302.42 (79.73)& 153.08 (46.53)\\ 
 &      & 54.56 & 79.76 & 46.64\\ \hline
    &PRS &217.53 (49.03) & 320.57 (68.79) &  162.36 (48.16)\\  
 & & 52.07 & 71.80 & 49.73  \\ \hline\hline
 $\delta=q_{1.00}$  & & & &\\\hline\hline
 &RT & 208.79 (28.09) & 308.29 (42.12) & 156.02 (28.45)\\ 
  & & {\bf  29.44} &  {\bf 42.92} &  {\bf 29.07} \\\hline
    &LT & 204.88 (54.63) & 302.43 (79.90) & 153.13 (46.79) \\ 
  &  &  54.85 & 79.93 & 46.89 \\  \hline
 & PRS  & 218.66 ( 51.33) & 321.70 (71.49) & 163.00 (50.13)\\ 
  & & 54.61 & 74.71 & 51.79\\ \hline
\end{tabular}
\end{center}
\caption{Simulation results for  a grid of 49 sites with spatial lag $k=1$. For each
estimation approach (RT, LT, PRS),  the mean and the standard errors (in brackets)  of the 
estimates are reported on the first line. RMSE values are given on the second line.}
\label{tab:49-1}
\end{table}

\begin{table}
\begin{center}
\begin{tabular}{|r|r|r|r|r|}\hline
   & Parameter            & $\sigma_{11}$ & $\sigma_{22}$ & $\sigma_{12}$ \\ \hline
   &True value  & 200.00 & 300.00 & 150.00 \\\hline\hline
   $\delta=q_{0.25}$  & & & &\\\hline\hline
    &RT & 204.41 (16.33) & 305.37 (25.63) & 152.06 (16.92)\\ 
 &  &  {\bf 16.91} &  {\bf 26.19} &  {\bf 17.04} \\ \hline
  &LT & 201.06 (35.39) & 302.03 (56.71) & 151.55 (32.59)\\ 
  & &35.41 & 56.74 & 32.63 \\\hline
 & PRS & 206.78 (29.79) & 312.02 (49.35) & 155.90 (31.49)\\ 
 &  & 30.55 & 50.79 & 32.04\\ \hline\hline
 $\delta=q_{0.50}$  & & & &\\\hline\hline  
  &RT & 205.58 (15.54) & 304.47 (24.01) & 148.33 (15.89)\\ 
  &  &  {\bf 16.51} &   {\bf 24.42} &  {\bf 15.97} \\ \hline
     &LT & 201.53 (35.92) & 302.52 (57.17)& 151.87 (33.07)\\ 
 &      & 35.95 & 57.23 & 33.13\\ \hline
    &PRS &208.22 (33.67) & 313.99 (54.34) & 156.79 (35.34)\\  
 & &  34.66 & 56.11 & 35.99  \\ \hline\hline
 $\delta=q_{1.00}$  & & & &\\\hline\hline
 &RT & 216.32 (13.80) & 306.31 (19.81) & 133.11 (13.26)\\ 
  & & {\bf 21.37}& {\bf 20.79} & {\bf  21.47 }\\\hline
    &LT & 202.20 (37.21) & 303.47  (58.34) & 152.31 (33.65) \\ 
  &  &  37.27 & 58.44 & 33.73 \\  \hline
 & PRS  & 211.34 (41.89) & 318.21 (64.93) & 158.78 (42.98)\\ 
  & & 43.40 & 67.43 & 43.87\\ \hline
\end{tabular}
\end{center}
\caption{Simulation results for  a grid of 49 sites with spatial lag $k=5$. For each
estimation approach (RT, LT, PRS),  the mean and the standard errors (in brackets)  of the 
estimates are reported on the first line. RMSE values are given on the second line.}
\label{tab:49-5}
\end{table}

\begin{table}
\begin{center}
\begin{tabular}{|r|r|r|r|r|}\hline
   & Parameter            & $\sigma_{11}$ & $\sigma_{22}$ & $\sigma_{12}$ \\ \hline
   &True value  & 200.00 & 300.00 & 150.00 \\\hline\hline
   $\delta=q_{0.25}$  & & & &\\\hline\hline
    &RT & 207.38 (7.24) & 299.57 (11.98) & 138.27 (7.98)\\ 
 &  &  {\bf 10.33 }&  {\bf 11.98} &  {\bf 14.18} \\ \hline
  &LT & 199.91 (18.79) & 300.53 (31.43) & 150.21 (18.07)\\ 
  & & 18.77 & 31.41 & 18.07 \\\hline
 & PRS & 201.55 (19.56) & 305.64 (32.64) & 152.30 (20.97)\\ 
 &  & 19.60 & 33.09 & 21.08\\ \hline\hline
 $\delta=q_{0.50}$  & & & &\\\hline\hline  
  &RT & 225.64 (6.59) & 309.70 (10.52) & 122.79 (7.06)\\ 
  &  & 26.47 &  {\bf 14.30} & 28.11 \\ \hline
     &LT & 200.16 (19.56) & 300.98 (32.61)& 150.39 (18.65)\\ 
 &      &  {\bf 19.56} & 32.61 &  {\bf 18.65}\\ \hline
    &PRS &202.43 (23.68) & 307.70 (39.42) & 153.23 (25.76)\\  
 & & 23.78 & 40.12 & 25.93  \\ \hline\hline
 $\delta=q_{1.00}$  & & & &\\\hline\hline
 &RT & 315.27 (7.45) & 397.33 (10.25) & 120.22 (7.19)\\ 
  & & 115.51 & 97.87 & 30.63 \\\hline
    &LT & 200.56 (22.75) & 302.21  (37.72) & 150.96 (21.70) \\ 
  &  &  {\bf 22.75} &  {\bf 37.75} &  {\bf 21.70} \\  \hline
 & PRS  & 204.09 (31.00) & 311.48 (52.11) & 155.00 (33.83)\\ 
  & & 31.24 & 53.31 & 34.17\\ \hline
  
\end{tabular}
\end{center}
\caption{Simulation results for  a grid of 49 sites with spatial lag $k=10$. For each
estimation approach (RT, LT, PRS),  the mean and the standard errors (in brackets)  of the 
estimates are reported on the first line. RMSE values are given on the second line.}
\label{tab:49-10}
\end{table}

\begin{center}
\begin{tabular}{|cc|rrr|rrr|}
\multicolumn{1}{c}{ \ } &\multicolumn{1}{c}{ \ } & \multicolumn{3}{c}{$k=5$}  &  \multicolumn{3}{c}{$k=10$} \\
\hline
                       &   &    0.9  & 0.95 &    0.98 &    0.9  & 0.95 &    0.98 \\
\hline
$\sigma_{11}$  & RT &50.48       & 30.65        & 16.32  &  226.47      &  167.81       &115.27      \\
\hline
                       & LT &0.80       &  1.55     &   2.20  & 0.06      &-0.04        &0.56         \\
\hline
$\sigma_{22}$  & RT &  40.03     & 19.27        &  6.31  &222.36      &155.91         & 97.33        \\
\hline
                       & LT & 0.99      & 1.89        &    3.47 &  0.18      &0.26         & 2.21      \\
\hline  $\sigma_{12}$  & RT & -16.48      & -18.61        &    -16.9   &   -8.3     &-20.17         &-29.78       \\
\hline
                       & LT &  0.65     &   1.33      &  2.31    &  0.10      &0.07         &0.96         \\\hline                    
\end{tabular}
\end{center}
\begin{table}
\caption{Bias values for the RT and LT estimates  using all pairs  and three different  threshold values, namely set to the 0.9, 0.95 and 0.98 empirical quantile. 
The spatial lags $k$ are  taken equal to $k=5$ and $k=10$.}
\label{tab:quantiles-distance}
\end{table}

\end{document}